%% file: paper.tex
\documentclass[conference]{IEEEtran}

\newcommand{\sectopic}[1]{\vspace{0.2em}\par\noindent{\textit{\bfseries #1}}}

\usepackage[utf8]{inputenc}
\usepackage{array}
\usepackage{booktabs}
\usepackage{multicol}
\usepackage{multirow}
\usepackage{tabularx}
\usepackage{longtable}
\usepackage{threeparttable}
\usepackage{enumitem} 
\usepackage{tcolorbox}
\usepackage{balance}
\usepackage{hyperref}
\usepackage{tcolorbox}
\usepackage{soul}
\usepackage[noadjust]{cite}

\newtcolorbox{myframe}[1][]{
  enhanced,
  arc=0pt,
  outer arc=0pt,
  colback=white,
  boxrule=0.8pt,
  #1
}
\newcommand{\rev}[1]{\textcolor{blue}{#1}}

\newcommand{\TBD}[1]{\textcolor{red}{#1}}
\newcommand{\crv}[1]{\textcolor{black}{#1}}

\begin{document}
\pagenumbering{arabic}

%\title{Model Generation from Requirements with LLMs:\\ the Case of UML Sequence Diagrams}

\title{Model Generation with LLMs:\\ From Requirements to UML Sequence Diagrams}

% \author{\IEEEauthorblockN{Anonymous author(s)}}


%\author{\IEEEauthorblockN{Anonymous author(s)}}
%\iffalse
\author{Alessio Ferrari\IEEEauthorrefmark{1},
\IEEEauthorblockN{Sallam Abualhaija\IEEEauthorrefmark{2},
Chetan Arora\IEEEauthorrefmark{3}
}
\IEEEauthorblockA{\IEEEauthorrefmark{1} Consiglio Nazionale delle Ricerche (CNR), Email: alessio.ferrari@isti.cnr.it}
\IEEEauthorblockA{\IEEEauthorrefmark{2} SnT University of Luxembourg, Luxembourg, Email: sallam.abualhaija@uni.lu}
\IEEEauthorblockA{\IEEEauthorrefmark{3} Monash University, Email: chetan.arora@monash.edu}
}
%\fi


%Telephone: (800) 555--1212, Fax: (888) 555--1212}


\maketitle
\begingroup\renewcommand\thefootnote{\textsection}
%\footnotetext{Corresponding Author}
\endgroup

% \thispagestyle{plain}
% \pagestyle{plain}
\begin{abstract}
\input{sections/abstract}

\end{abstract}


% \fi
\IEEEpeerreviewmaketitle

\input{sections/introduction}
\input{sections/background}

\input{sections/design}

\input{sections/results}
\input{sections/discussion}
\input{sections/threats}
\input{sections/conclusion}
%\input{sections/acknowledgement}


%\section*{Acknowledgment}




%\bibliographystyle{IEEEtran}
%\newpage

\bibliographystyle{IEEEtran}
\balance

\bibliography{bibliography}

\end{document}

%% file: sections/abstract.tex
Complementing natural language (NL) requirements with graphical models can improve stakeholders’ communication and provide directions for system design. However, creating models from requirements involves manual effort. The advent of generative large language models (LLMs), ChatGPT being a notable example, offers promising avenues for automated assistance in model generation. This paper investigates the capability of ChatGPT to generate a specific type of model, i.e., UML \textit{sequence diagrams}, from NL requirements. We conduct a qualitative study in which we examine the sequence diagrams generated by ChatGPT for 28 requirements documents of various types and from different domains. 
%Our study aims to uncover potential  issues that emerge in the models generated by ChatGPT, thereby hindering its applicability in practice. 
Observations from the analysis of the generated diagrams have systematically been captured through evaluation logs, and categorized through thematic analysis. Our results indicate that, although the models generally conform to the standard and exhibit a reasonable level of understandability, their completeness and correctness with respect to the specified requirements often present challenges. This issue is particularly pronounced in the presence of requirements smells, such as ambiguity and inconsistency. 
%Our results show that, while the models tend to adhere to the standard and are generally understandable, their correctness with respect to the requirements is frequently an issue
%\rev{according to the requirements types} 
%and 
%is highly impacted by requirements smells, such as ambiguity and inconsistency. 
The insights derived from this study can influence the practical utilization of LLMs in the RE process, and open the door to novel RE-specific prompting strategies targeting effective model generation.

%% file: sections/introduction.tex
\section{Introduction}
\label{sec:introduction}

% \hl{Story outline: 
% - Models are effective. 
% - Generating models from scratch is laborious.
% - UML is the most widely used representation for models. 
% - Exsiting work provides some automated support based on X and Y approaches (ideally also using NLP). 
% - Such approaches have limitations. 
% - Recent NLP technologies, ChatGPT, offer promising avenues for automating model generation. 
% - LLMs are known to introduce false information and plausible answers. 
% - There is a need for idenitfying the various issues emerging from utilizing LLMs for model generation. 
% }
Graphical models are recognized to be an effective tool for facilitating communication between different stakeholders involved in the requirements engineering (RE) process and guiding towards the design of a system~\cite{jolak2020software}. 
%,bork2021technique}. 
However, requirements are typically written in natural language (NL)~\cite{wagner2019status}, and complementing them with models requires significant manual effort~\cite{ambriola2006systematic}. The support of natural language processing (NLP) tools for model generation can greatly facilitate the work of requirements engineers and streamline the RE processes~\cite{arora2019active}.
%{yue2011systematic,
%Several contributions in the literature have focused on generating models in the form of Unified Modelling Language (UML) diagrams~\cite{uml}. 

%\TBD{RE models can adopt various graphical notations, including well-known goal-oriented RE notations such as KAOS~\cite{van2001goal}, i*~\cite{yu1997towards}, and User Requirements Notation (URN)~\cite{amyot2003introduction},} 
RE models can use various graphical notations, including well-known goal-oriented RE notations~\cite{horkoff2019goal}, along with commonly adopted representations such as Unified Modeling Language (UML) diagrams~\cite{uml}. 
UML is a widely known semi-formal language for software design and requirements modelling~\cite{wagner2019status}, which includes structural models, e.g., class diagrams, and behavioral ones, e.g., sequence diagrams. 

%Previous studies concerning UML model generation include the CIRCE requirements analysis environment~\cite{ambriola2006systematic}, aToucan~\cite{yue2015atoucan}, a tool for generating class, sequence, and activity diagrams, and the more recent proposal by Jahan \textit{et al.}~\cite{jahan2021generating}. The majority of these contributions
%Previous contributions concerning UML model generation (e.g., \cite{ambriola2006systematic,yue2015atoucan}) 
%Previous contributions concerning UML model generation (e.g., \cite{ambriola2006systematic,yue2015atoucan}) 
%require controlled natural language formats as input rather than unrestricted textual requirements---which are more common in practice~\cite{wagner2019status}---and tend to use heuristic rule-based approaches~\cite{ahmed2022automatic}. 

Existing work in RE for automatically generating UML diagrams from  %targeting UML model generation from 
requirements typically use heuristic rule-based NLP approaches~\cite{ambriola2006systematic,yue2015atoucan,jahan2021generating}.
Such approaches have several limitations, %pose several issues, 
including 
%, including limited scalability, 
significant manual effort for construction and maintenance, and difficult adaptability to different contexts~\cite{saini2022automated}. 
With the recent advances in NLP technologies in general and generative large language models (LLMs) in particular, some of these limitations can now be overcome. 
%it becomes possible to overcome some of these limitations
~\cite{Arora:23}. 
%Generative 
LLMs 
%are typically pre-trained---which reduces end-user effort---,
exhibit acceptable contextual understanding, are typically pre-trained, 
%~\cite{min2023recent}, 
and can be used out-of-the-box, thus reducing human effort in building a model generation tool. 
%Given that model definition requires creativity, and that LLMs appear to be able to emulate this desirable quality~\cite{}, it is important to investigate what are the potentials of LLMs for this task. 

%However, LLMs are typically designed to provide plausible answers rather than \textit{right} answers and can be prone to hallucinations~\cite{fan2023large}. Before integrating them into the RE process, it is essential to identify the potential problems a requirements engineer might face when using them for model generation. 
%Possible issues include quality issues in the generated models, such as incompleteness, incorrectness, or poor understandability. 

%  These are behavioral models that represent interactions between different components of the system-to-be in terms of function calls and messages. They are particularly useful, as they are able to represent the dynamic behavior of a system, which is complementary to the structural (static) view given by the class diagrams. 
%, where problems are intended as undesired characteristics of the produced diagrams, such as incompleteness, incorrectness, or poor understandability.

%While previous exploratory work evaluating the issues of using LLMs for model generation considers goal models~\cite{chen2023use} or UML class diagrams~\cite{camara2023assessment,chen2023automated}, 

%This paper aims to investigate
%possible emerging issues when generating UML \textit{sequence diagrams} using ChatGPT, a well-known LLM. 
This paper aims to examine the capability of ChatGPT, a well-known LLM, to generate UML \textit{sequence diagrams}. 
Sequence diagrams are behavioral models that represent interactions between different components of the system-to-be~\cite{uml}.  
%in terms of function calls and messages. 
They are particularly useful, as they are able to represent the dynamic behavior of a system, which is complementary to the structural (static) view given by class diagrams. Our motivation for focusing on sequence diagrams is that, despite being
arguably simple to understand by different stakeholders, they are less studied in the RE literature~\cite{ahmed2022automatic}. Previous studies exploring the problems of using LLMs for model generation consider goal models~\cite{chen2023use} or class diagrams~\cite{camara2023assessment,chen2023automated}, but, to our knowledge,  
none of the studies target sequence diagrams.

%To identify the issues emerging in sequence diagram generation, 
To analyze the model generation capability of ChatGPT, we performed an exploratory study\footnote{\crv{The study is a hybrid between a judgment study and a sample study~\cite{stol2020guidelines}, cf. Sect.~\ref{sec:discussion} for a discussion on the study design.}} involving three researchers---the authors of this paper---who have 7 to 13 years of experience in both RE and NLP, and practical confidence with sequence diagrams. 
Specifically, two researchers prompted ChatGPT to generate sequence diagrams corresponding to 28 NL requirements documents covering requirements specified in different formats, including ``shall''-style requirements, user stories, and use case specifications 
(\crv{cf. Sect.~\ref{sec:sequencediagram} and ~\ref{sec:rq2}  for examples of requirements-diagram pairs}). 
%Two of them considered a set of \hl{37} NL requirements documents in different formats, namely ``shall'' requirements, user stories, and use case specifications, and prompted ChatGPT to generate models.
%To uncover the potential issues in LLM-based model generation, 
The researchers also introduced variants of the same requirements to simulate realistic scenarios, hence exposing ChatGPT to common challenges. 
%\TBD{also considered variants of the original requirements, which were aimed to expose the LLM to realistic challenges, such as}
Examples of challenges include the presence of smells or the evolution of requirements, i.e., addition, removal, modifications. The researchers \textit{scored} the quality of the diagrams according to different criteria and tracked their observations in structured evaluation logs, one for each generated diagram. The observations were oriented to highlight quality \textit{issues} in the diagrams. Following this, the third researcher performed a thematic analysis~\cite{guest2011applied} on the evaluation logs and identified 23 main categories of issues. From the results, it emerges that the generated diagrams score well for understandability, standard compliance, and terminological alignment with the requirements. However, they exhibit significant issues related to completeness and correctness, such as missing/incorrect elements, structural issues, or components that deviate from what is specified by the requirements. These issues become more evident in the presence of low-quality requirements that include ambiguities or inconsistencies, and when technical/contextual knowledge is needed to interpret the requirements.  
%particularly in the presence of low-quality requirements. 
%that the diagrams often offer an incorrect representation of the requirements, despite their apparent clarity, and their quality is highly affected by the quality of the source requirements. 

\sectopic{Contributions. }
Our study contributes with a structured framework of issues associated with the generation of sequence diagrams from NL requirements through ChatGPT. Our results outline possible avenues for future research. These include the need for iterative, RE-specific prompting solutions, as well as the need to address tacit/domain knowledge issues that affect a general-purpose LLM such as ChatGPT, when dealing with technical requirements data. 
 %\hl{list of topics from the discussion}. 

%\sectopic{Structure. } The remainder of the paper is structured as follows. Section~\ref{sec:background} presents the background and reviews the related work. Section~\ref{sec:design} describes our design method. Section~\ref{sec:results} discusses the results of our study. Section~\ref{sec:discussion} provides further insights derived from our study. Section~\ref{sec:threats} outlines the validation considerations, and Section~\ref{sec:conclusion} concludes the paper.     

%% file: sections/background.tex
\section{Background and Related Work}\label{sec:background}
%This section presents the background for our study and further reviews the related work.  
\subsection{Sequence Diagrams}
\label{sec:sequencediagram}
\begin{figure} [t!]
\centering
\includegraphics[width=0.41\textwidth]{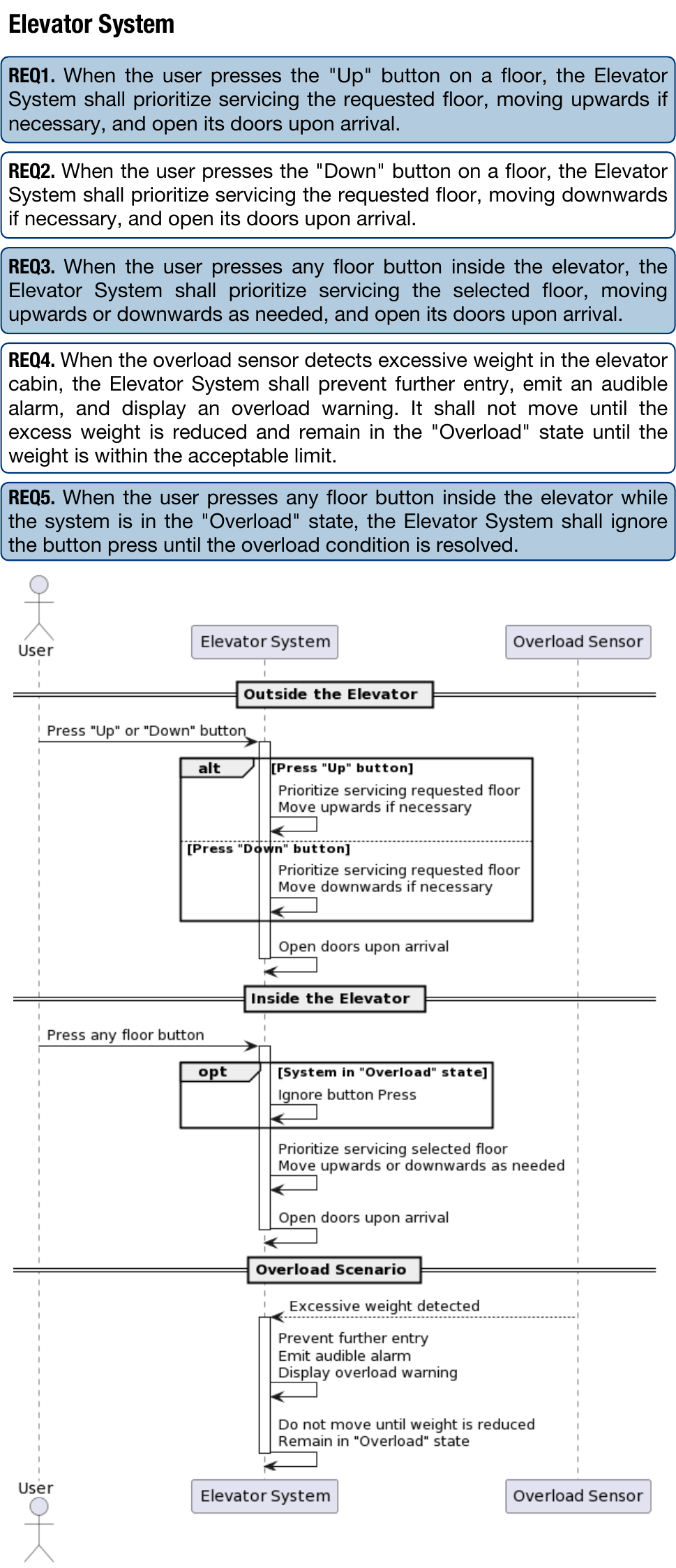}
\vspace{-.5em}
\caption{Example requirements for an ``elevator system'' and the corresponding sequence diagram.}
\label{fig:example}
\vspace{-1.5em}
\end{figure}

%\sectopic{Sequence Diagrams. } 
UML sequence diagrams are models %graphical models that 
representing interactions between different system components in terms of function calls and messages~\cite{uml}. Fig.~\ref{fig:example} shows an example of 
%``shall'' 
requirements for an elevator system and the corresponding sequence diagram. The user is represented with a stylized figure, while system components are represented as rectangles at the top and at the bottom of the diagram. In the figure, we see the User and two components: the Elevator System and the Overload Sensor. Components are associated with vertical lifelines,
%---vertical rectangles indicate when a component is active---
while the horizontal arrows identify function calls (solid line) and messages (dashed line). The models can also include alternative choices (\textbf{alt}, e.g., Press ``Up'' (``Down'') button), and optional steps (\textbf{opt}, e.g., System in ``Overload'' state), enclosed in boxes. The syntax also allows separation into conceptual blocks, e.g., ``Outside the Elevator'', identified by horizontal lines. The given example is a highly simplified case, which we use to introduce the syntax of sequence diagrams. In our study, we use more complex, real-world requirements cases. 
%The requirements used in the study are more complex, and the figures in Section V show only a portion of the requirements used for diagram generation, as well as a zoom of the diagrams. 
%Sequence diagrams for the other types of requirements, also for an elevator system but capturing different aspects, are represented in \hl{Fig}~\ref{fig:}. We see that the different 
%\rev{Other requirements types (e.g., user stories) introduce different degrees of abstraction of the requirements that impact the generated diagrams. We, therefore, scrutinize diagrams generated from different requirements types.} 
We note that contrary to formal sequence diagrams, where the labels on the arrows include pseudo-code, we generate abstract diagrams, where the labels are free-form text. The reason is that our analysis aims to explore models for facilitating interaction with stakeholders who may not have programming experience. 

\input{sections/related}

%% file: sections/related.tex
\subsection{Related Work}\label{related-work} 

%This section reviews the related work
%on model generation and the application of LLMs in RE. We 
%and outlines the research gap that our study aims to address.  
\sectopic{Model Generation.} 
Model generation is a key task in NLP for RE. This 
involves creating model abstractions---typically in a graphical form---from input requirements.  %that can be expressed in different formats, e.g., ``shall'' requirements, user stories, \textit{etc.} 
The systematic mapping study from Zhao \textit{et al.}~\cite{zhao2021natural}, which analyses studies in NLP for RE from 1983 until 2019, 
reports 59 contributions on model generation, with the majority of automated NLP tools targeting this task. 
%Model generation is a key task in NLP for RE~\cite{zhao2021natural}. 
%This 
%involves creating model abstractions---typically in a graphical form---from input requirements.  %that can be expressed in different formats, e.g., ``shall'' requirements, user stories, \textit{etc.} 
%The systematic mapping study from Zhao \textit{et al.}~\cite{zhao2021natural}, which analyses studies in NLP for RE from 1983 until 2019, 
%reports 59 contributions on model generation, with the majority of automated NLP tools targeting this task.  
%and the majority of the identified tools (26\%) target this task. 
%Model generation can take different flavors~\cite{zhao2021natural}, from the generation of models to support requirements elicitation, analysis, and design~\cite{yue2015atoucan,chen2023use}, to the synthesis of feature models in a product-line engineering context~\cite{becan2016breathing}, to the generation of high-level models 
%of early requirements or user stories 
%to support project scoping~\cite{lucassen2017extracting}. 
%review the main reference works concerning UML model generation since this is the main focus of our contribution. 
%For more details, we refer the reader to 
%Ahmed \textit{et al.} provide a review on the topic~\cite{ahmed2022automatic}.
%In the following, we focus on UML model generation, which is most pertinent to our study.
%We focus below on UML model generation which is the most relevant to our work. 
%The majority of the works focus on 
Model generation can take different flavors, from the generation of models to support requirements elicitation, analysis, and design, to the synthesis of feature models in a product-line engineering context, to the generation of high-level models of early requirements~\cite{zhao2021natural}.
%or user stories to support project scoping~\cite{lucassen2017extracting}. 
%review the main reference works concerning UML model generation since this is the main focus of our contribution. 
For more details, we refer the reader to a recent comprehensive review~\cite{ahmed2022automatic}.
In the following, we focus on UML model generation, which is most pertinent to our study.

The RE community has extensively investigated UML  \textit{class diagram} generation~\cite{saini2022automated,arora2019active}, typically used for representing domain models (i.e., high-level abstraction of domain entities and their relations). 
Generating \textit{sequence diagrams} or other behavioral representations has also been investigated in RE to some extent. 
%Some of the existing contributions also address the generation of \textit{sequence diagrams} or similar behavioral representations such as message sequence charts (MSCs). 
%Among them, the early contribution of 
Kof~\cite{kof2007scenarios} generates message sequence charts (MSCs) from scenario descriptions using a rule-based approach. 
%, which can also identify missing objects and actions in the original descriptions. 
%The approach is evaluated on an industrial specification from the automotive domain. 
Yue \textit{et al.}~\cite{yue2015atoucan} present aToucan, a tool that extracts sequence diagrams from use case specifications expressed in a constrained natural language. The tool relies on %is based on 
%a set of parsing and transformation 
parsing/transformation rules and uses an intermediate representation to pass from the specifications to the final models. %aToucan is evaluated on two industrial case studies at Cisco Systems, Inc. 
More recently, Jahan \textit{et al.}~\cite{jahan2021generating} 
present a rule-based approach that, 
%which, 
unlike the other works, 
can be applied to free text use case specifications. 
%The approach is evaluated on three simple problems.
%that goes beyond previous contributions by accepting free text use case specifications instead of constrained natural language ones. Similarly to the other works, the approach is rule-based and evaluated on three simple problems.

%Other contributions focused on sequence diagrams and based on rules, but with limited evaluation, are, e.g., Thakur and Gupta~\cite{} and  

\sectopic{Large Language Models in RE.} 
% LLMs are deep neural networks that use a transformer architecture and are pre-trained on large amounts of text via self-supervised learning to capture the statistical regularities and semantic knowledge in natural language~\cite{nguyen2023generative,min2023recent}. They are typically used as \textit{generative} models to produce human-like responses to natural language queries. LLMs can address several NLP tasks, such as question-answering, summarization, and translation. In our paper, we focus on question answering, a task consisting of generating the response to a question, normally referred to as \textit{prompt}. Well-known examples of LLMs for question answering are ChatGPT, Bing Chat (based on GPT4), Claude 2, Bard, and Vicuna (based on Llama). 
 LLMs are deep neural networks that use a transformer architecture and are pre-trained on large amounts of text via self-supervised learning to capture the statistical regularities and semantic knowledge in natural language~\cite{min2023recent}. Recent LLMs are \textit{generative} models that produce human-like responses to NL queries (also referred to as \textit{prompts}). 
Existing surveys acknowledge that the application of LLMs in RE is particularly scarce compared to other software engineering areas, such as testing, code generation, and program repair
~\cite{fan2023large,hou2023large}. %In the following, we report some relevant examples. 
LLMs have been investigated in several contexts in RE, e.g., summarization from legal texts~\cite{jain2023transformer} and requirements traceability~\cite{rodriguez2023prompts}. 
\iffalse
Jain \textit{et al.}~\cite{jain2023transformer} investigate utilizing LLMs for requirements summarization from legal texts. 
They first generate summaries using the GPT-3 model and then use these auto-generated summaries for fine-tuning Pegasus, GPT-2, and other open-source models. % that do not require the disclosure of data, as in the case of GPT-3. Then, 
The authors evaluate the performance of LLMs on industrial datasets and report Pegasus as the best model for their task. %outperforms the other options. 
Marczak \textit{et al.}~\cite{marczak2023using} use ChatGPT to generate user stories reflecting human values, which are then used as creativity triggers in requirements elicitation sessions. The evaluation with focus groups shows that the generated user stories effectively inspire participants to identify value-relevant requirements. Rodriguez \textit{et al.}~\cite{rodriguez2023prompts} experiment with different prompting strategies for requirements traceability. % with the aid of Anthropic’s Claude instant model. 
Their evaluation %exploratory evaluation with the CM1 dataset 
shows that slightly varying prompts can significantly impact the output. %lead to significant differences in the outputs. 
\fi
%
More relevant to our work, %Some works also explore the potential of LLMs for model generation. Among them, 
Chen \textit{et al.}~\cite{chen2023use} evaluate the potential of GPT-4 for generating goal models using the textual grammar for the Goal-oriented Requirement Language (GRL) based on NL descriptions of the problem context. 
%Their experiments 
%on four example cases 
%show that incremental prompts can improve the results, and domain knowledge is highly needed to assess the plausible---yet frequently incorrect---output. 
Chen \textit{et al.}~\cite{chen2023automated} evaluate GPT-3.5 and GPT-4 for generating class diagrams from NL descriptions. 
%Their evaluation shows that %on a ground truth from an undergraduate course 
%generating relationships between classes is challenging for LLMs. %shows that the LLMs struggle to . 
%Furthermore, they show that while few-shot approaches (i.e., adding examples to the prompts) can improve performance, chain-of-thought prompting (i.e., adding reasoning steps) leads to a performance decrease. Still on class diagram generation, 
Camára \textit{et al.}~\cite{camara2023assessment} conduct an exploratory study on using ChatGPT for class diagram generation. %They exercise the model with different sample problems and provide some lessons learned. 
The authors conclude that iterations are needed to produce models of sufficient quality. 
%They further observe that the adherence to the standard of the generated models tends to be poor. 
Other studies in RE propose pattern catalogs of prompts for specific problems, such as classification and traceability~\cite{ronanki2023requirements,white2023chatgpt}. 

\sectopic{Research Gap.} 
Most of the studies 
%discussed above 
on model generation use rule-based approaches~\cite{ahmed2022automatic}. Such approaches 
%These typically 
require defining a complex set of heuristic rules, which are hardly maintainable and poorly adaptable.  %to different contexts. 
%One notable exception is the contribution by 
Saini \textit{et al.}'s approach for class diagram generation is an exception~\cite{saini2022automated} in that it combines rules with machine learning. However, their approach does not exploit LLM capabilities. 
%but is based on more traditional solutions.  
%full potential of the recent NLP landscape dominated by LLMs. %  However, this work does not use the most recent LLM-based technologies. 
Current works on model generation using LLMs~\cite{camara2023assessment,chen2023automated,chen2023use} do not focus on sequence diagrams and %have limited generalizability since 
they mainly use toy requirements instead of real-world specifications. 
% The studies using LLMs in RE are scarce, but some focus on model generation. These are the exploratory studies by Camara \textit{et al.}~\cite{camara2023assessment} and Chen \textit{et al.}~\cite{chen2023automated}, targeting UML class diagram synthesis, and Chen et al.~\cite{chen2023use}, targeting goal models. However, these works do not focus on sequence diagrams and consider sample prompts rather than real-world specifications. 
In contrast to existing work, our study is,
to the best of our knowledge, the first to: (1) target UML sequence diagram generation using LLMs; (2) consider industrial requirements specifications as input, belonging to different domains and having different formats.  
%(3) consider different requirements types, namely ``shall'' requirements, user stories, and use case specifications.

%% file: sections/design.tex
\section{Research Design}
\label{sec:design}

The overarching goal of our study is to examine the capability of ChatGPT to generate sequence diagrams. While sequence diagrams can serve various purposes, including code generation~\cite{kundu2013automatic}, our study specifically analyzes their role in \textit{complementing} requirements, aiming to facilitate communication with stakeholders. 
%primarily to complement the requirements with the goal of facilitating communication among stakeholders.
%as highlighted in Section~\ref{sec:introduction}. 
%Therefore, the automatically generated sequence diagrams were assessed, among others, for whether they  \textit{complement} the requirements to facilitate system design. 
%
Our study is guided by 
the following research questions (RQs): 

\noindent\textbf{RQ1:} \textit{What is the quality of the sequence diagrams generated from NL requirements by ChatGPT?}
%\textit{How reliable is ChatGPT for generating sequence diagrams from NL requirements?} 
RQ1 aims to provide a \textit{quantitative} evaluation of the quality degree of the diagrams generated by ChatGPT, thus giving an indication of its applicability in practical settings.

\noindent\textbf{RQ2:} \textit{What are the issues emerging from using ChatGPT for generating sequence diagrams from NL requirements?}
RQ2 aims to \textit{qualitatively} explore the problem domain and produce a catalog of typical issues (e.g., incompleteness, low level of understandability) that can emerge when generating sequence diagrams using ChatGPT. By \textit{issue}, we intend any observable problem in the generated diagram, possibly associated with some specific characteristic of the input requirements or limitations of ChatGPT.

We perform our study on ChatGPT based on the GPT3.5 model available through the web application\footnote{\url{https://chat.openai.com/}}. 
Our rationale behind selecting GPT3.5 is that is free to use and provides an intuitive interface, which can be used by analysts who do not have the coding skills required to use the OpenAI API. 

\subsection{Data Collection}\label{subsec:data-collection}
\input{sections/data-collection}

\subsection{Data Analysis}
%\hl{I am now assuming we have two RQ, a quantitative and a qualitative one.}

For \textbf{RQ1}, we first assessed that A1 and A2 had similar interpretations of the score values according to the scale. To this end, an independent cross-evaluation was performed. A1 inspected and scored 15 of the diagrams produced by A2, and vice versa---a total of 30 models were cross-evaluated. 
The agreement between A1 and A2, computed through a square-weighted Cohen's Kappa~\cite{Cohen:60}, led to $\kappa=$ 0.67, indicating substantial agreement. With square-weighted Kappa, disagreements are weighted according to their squared distance from perfect agreement, thus penalizing larger disagreements in the scale.  
%We used a weighted formula to account for small differences in the scoring, i.e., cases in which the scores differed only by one were considered to indicate agreement. 
We further used the nonparametric Wilcoxon signed rank test for each criterion to check whether the average scores significantly differed from the mean value of 3, as done e.g., in~\cite{abrahao2011evaluating}, thus suggesting a high degree of fulfillment of the criterion. Specifically, we tested the null hypothesis: \textit{The scores for [criterion] do not differ from the mean value}, considering $\alpha = 0.05$. We also evaluated the effect size with Cohen's d~\cite{cohen2013statistical}. For this evaluation, we used solely cases that did not include modifications or smells. The reason is that such alterations could lead to inaccurate diagrams, while here we want to check the reliability of ChatGPT starting from well-formed requirements.

%# Install and load the 'irr' package
%install.packages("irr")
%library(irr)

%# Create a data frame with the ratings from two raters
%data <- data.frame(
%  Rater1 = c(3, 4, 2, 5, 3),
%  Rater2 = c(3, 4, 2, 5, 2)
%)

%# Define weights for agreement within one scale point
%weights <- matrix(c(0, 1, 1, 1, 0), ncol = 5)

%# Calculate Weighted Kappa
%weighted_kappa_result <- kappa2(data, "weighted", weights = weights)

%# Print the result
%print(weighted_kappa_result)

For \textbf{RQ2}, an author of the paper not involved in the evaluation (A3) performed the thematic analysis according to Clarke and Braun~\cite{clarke2017thematic} through semi-open coding in NVivo, using the logs produced during the data collection phase. The analysis aimed to identify and classify issues encountered during the generation of sequence diagrams. Closed codes are the criteria (completeness, correctness, etc.) that we regard as high-level (HL) codes (our pre-defined meta-level codes based on the five criteria discussed earlier). For each HL code, open coding was performed. A3 went through the logs and annotated them with low-level (LL) codes, asking for clarifications when some of the logs were not entirely understandable. For example, the statement ``Failure management is also missing as a component'' was coded as \textit{Missing Component}, under the HL code \textit{Completeness}. 135 LL codes were introduced at this stage, partitioned into different HL ones, plus an additional one, \textit{General Issues}, including LL codes that could not be directly associated with the evaluation criteria. Then, the LL codes were revised by A1, who also inspected the associated logs and suggested removing or merging some of the codes. This sanitization process resulted in 62 LL codes, which were then aggregated into 23 mid-level (ML) codes---still grouped according to the HL codes, plus \textit{General Issues}. For instance, \textit{Missing Component} and \textit{Missing Condition} were grouped under the ML code \textit{Summarization Issues}, under the HL code \textit{Completeness}. A2 finally inspected the results of the thematic analysis. The involvement of three subjects was aimed at mitigating the inherent subjectivity of thematic analysis. 
In defining the codes, we did not differentiate among requirements formats. %\crv{By not differentiating among different requirements formats, our analysis may have overlooked some format-specific issues and produced a generalized understanding of issues. However, we note that our primary focus in this study was on building a general understanding of the efficacy of LLMs in generating models from requirements.}
Furthermore, we did not systematically trace requirements modifications with issues. The analysis of format-related aspects and the correlation between issues and modifications requires a more systematic analysis, which is outside the scope of this study. Here, we are mainly interested in general common issues that can inspire research questions for future investigations.

%% file: sections/data-collection.tex
%\hl{old and short}

\sectopic{Datasets.} Our data collection aimed at manually examining the sequence diagrams generated by ChatGPT and identifying issues that affect the quality of these diagrams. 
%and hence hinder their integration in the RE process. 
To achieve our goal, we collected 28 industrial requirements documents covering diverse application domains. The documents originate from three sources: (i) the ``Ten Lockheed Martin Cyber-Physical Challenges''\footnote{\url{https://github.com/hbourbouh/lm_challenges}} containing ten requirements documents from the cyber-physical domain.  
%inspired by flight control and vehicle management systems.  %and have been used in the literature, e.g., by Mavridou \textit{et al.}~\cite{mavridou2020ten}. 
(ii) The PURE dataset~\cite{ferrari2017pure}, containing 79 documents that cover multiple application domains and requirements formats (e.g., ``shall'' requirements, use case specifications). 
%The PURE dataset has been leveraged for solving various RE tasks, e.g., referential ambiguity~\cite{ezzini2022automated}. 
(ii) A dataset of user stories~\cite{dalpiaz2020conceptualizing}.
%, previously used for conceptual modeling. 
%This dataset has been released as part of conceptual modeling efforts from requirements.  
%namely the PURE dataset~\cite{?}, the Lockheed Martin challenges~\cite{}, a publicly available dataset of user stories~\cite{}, and a set of 10 sample requirements provided by the authors.

The criteria for selecting the documents are:\\ 
(i) \textit{Diversity in structure, domain, and requirements types:} The documents cover diverse domains (18 in total, cf. Table~\ref{tab:data-col}), 
%such as healthcare, security, and railway, 
as well as different requirements types, namely ``shall'' type 
%(s in Table~\ref{tab:data-col}), 
user stories, and use case specifications. \\
(ii) \textit{Representativeness of real software projects:} Unlike toy requirements, the selected documents contain requirements from real industry projects.

\sectopic{Variants.} From each of the 28 selected documents, two authors of this paper (A1 and A2) extracted one requirements subset that was considered amenable to be represented as a sequence diagram, and generated a model from it. 
%To draw meaningful conclusions, 
In addition, A1 and A2 explicitly challenged ChatGPT by manually introducing a set of variants for each requirements subset. 
These variants encompass the intentional introduction of specific changes in the requirements, such as modifying, adding, or deleting a requirement. Additionally, the variants could introduce smells expected to occur in the requirements, e.g., ambiguity, inconsistency, and incompleteness. Introducing such changes exposes the LLMs to practical challenges in the RE
field, i.e., the evolution of requirements throughout the project lifetime and the presence of smells. 
We selected only functional requirements, as sequence diagrams are more appropriate to represent behavior rather than quality aspects. %Introducing such changes in the requirements exposes the LLMs to realistic challenges in the RE field, i.e., the evolution of requirements throughout the project lifetime and the presence of smells.
%The variant can involve injecting a particular (and meaningful) change in the requirement, e.g., modifying a requirement, adding a new requirement, or deleting a requirement. 
%The requirements ambiguities, duplications, and incomplete information.
%The variant can also involve gluing together different subsets of the requirements documents. 
%Defining the variants was performed independently in a semi-systematic manner. Specifically, A1 and A2---who have substantial experience in RE and quality---incrementally defined the variants in a greedy-like manner according to potential issues expected to occur in requirements. 
%Defining the variants was performed independently in a greedy-like manner. Specifically, 
A1 and A2---who have 7 to 13 years experience in quality requirements and NLP for RE, and practical confidence with sequence diagrams---incrementally defined the variants in a \textit{greedy-like} manner, according to their intuition of what could be a realistic modification. 
%according to potential realistic changes expected to occur in the requirements according to their experience. 
%Introducing such changes in the requirements 
%to expose the LLM to realistic challenges in the RE field, i.e., the evolution of requirements throughout the project lifetime and the presence of smells. 
%For example, requirements can contain various ambiguity types, duplications, or incomplete information. 
%\rev{While a more systematic generation of variants (i.e., introducing all possible changes in a given document) would have allowed a more complete exploration of the problem space, However, it was not considered feasible and would have constrained the creativity of the authors in identifying issues. 
A systematic generation of variants (i.e., introducing all possible changes in a given document) would have allowed a more complete exploration of the problem space. However, it was considered hardly feasible and would have constrained the creativity of the authors in introducing realistic changes. 
%Introducing such changes exposes the LLMs to practical challenges in the RE
%field, i.e., the evolution of requirements throughout the project lifetime and the presence of smells. 

Information about the requirements files and number of requirements in each requirements subset (REQ) is reported in Table~\ref{tab:data-col}. 
%to cover different domains (\hl{X domains}) and requirements types. 
A1 considered 44 variants of 19 subsets (the term ``variant'' also includes the original subset), and A2 considered 43 variants of 12 subsets---3 were common, and initially used to ensure alignment in the evaluation strategy. 
%for initial alignment in terms of evaluation approach. 
Our data collection resulted in analyzing 17 subsets of ``shall'' requirements (57  variants), 7 use case specifications (17 variants), and 4 user story documents (13 variants). In total, 87 variants were produced. 
These different numbers reflect the order of evaluation (i.e., ``shall''  requirements were evaluated first), as A1 and A2 incrementally selected the requirements documents and interrupted their evaluation when no additional issues emerged, i.e., a form of saturation~\cite{braun2021saturate} was reached. 
%In addition, fewer issues emerged for use cases and user stories despite the challenging changes. 
We acknowledge a bias due to this order, which could not be entirely mitigated.

\input{sections/tab-collection}

\sectopic{Diagram Generation.} We generated the sequence diagram for each variant %requirements subset and variant thereof 
%and variant thereof  %document and mutant thereof (explained later below), 
by prompting ChatGPT using the \textit{visualization generator pattern}~\cite{white2023prompt}: ``Generate a sequence diagram from these requirements so that I can provide it to Planttext to visualize
it. Requirements: \{list of requirements\}''. 
We selected Planttext\footnote{\url{https://www.planttext.com}/} to visualize the sequence diagrams as it is web-based, free- and easy-to-use, and it applies the PlantUML textual language~\footnote{\url{https://plantuml.com/guide}}, a widely-used and human-readable language. In our analysis, we used solely the diagrams resulting from this prompt (i.e., no iterations were performed, and we used a separate session for each prompt). %While sometimes we further prompted ChatGPT for corrections, the diagrams resulting from these iterations are not evaluated in this study. 

%Overall, 28 documents were selected 
%to cover different domains and requirements types. A1 considered 44 variants of 19 documents, and A2 considered 43 variants of 12 documents. 
\sectopic{Evaluation.} For each generated diagram, A1 and A2 independently performed a critical evaluation, which was documented in a textual log file. The evaluation was performed according to the following quality criteria. 

%Overall, A1 and A2 considered 18 documents of “shall” requirements (60 logs), 6 use case specifications (19 logs), and 4 user story documents (8 logs).  These different numbers reflect the order of evaluation (i.e., ``shall''  requirements were evaluated first), as A1 and A2 interrupted their evaluation when no additional issues emerged, i.e., a form of saturation~\cite{braun2021saturate} was reached. In addition, fewer issues emerged for use cases and user stories despite the challenging mutations. We acknowledge a bias due to this order, which could not be entirely mitigated.

%such a systematic method is infeasible and does not reflect a real scenario in practice.}   

%Following this, A1 and A2 generate multiple sequence diagrams for each document. Finally, A1 and A2 evaluate the resulting sequence diagrams according to the following quality criteria.  

\begin{itemize}[wide]
    \item \textbf{Completeness:} The diagram covers the content of all the requirements (external completeness~\cite{zowghi2003interplay}) with a sufficient degree of detail to communicate with potential stakeholders.
    \item \textbf{Correctness:} The diagram specifies a behavior that is coherent and consistent with the requirements.  %the diagram is consistent
    %and possible abstracted information does not compromise its adherence to the requirements. 
    \item \textbf{Adherence to the standard:} The diagram is syntactically correct (i.e., it can be interpreted by PlanText\footnote{PlantText may not be fully compliant with the UML standard, but we take its syntax as a reference to avoid manual checking of all the nuances of the reference standard.}) and semantically sound (i.e., it uses constructs appropriately). 
    \item \textbf{Degree of understandability:} The diagram is sufficiently clear, given the complexity of the requirements, and does not contain redundancies. 
    \item \textbf{Terminological alignment:} The terminology used in the generated diagram aligns with the one in the requirements. 
\end{itemize}

The criteria were established by the authors through consensus, drawing from preliminary experiments oriented to identify relevant quality dimensions, and considering those criteria for requirements sets from ISO/IEC/IEEE 29148:2018(E)~\cite{8559686}
%outlined by Montgomery \textit{et al.}~\cite{montgomery2022empirical} 
that were considered applicable to models. 

Each criterion was assessed according to a five-point ordinal scale 
%Likert scale~\cite{likert1932technique}, 
where each integer indicates a degree of fulfillment of the criterion 1 = ``Not fulfilled at all''; 2 = ``Fulfilled to a minimal extent'', 3 = ``Partially fulfilled'', 4 = ``Mainly fulfilled''  5 = ``Completely fulfilled''. This information was then analyzed to answer \textbf{RQ1}. For each criterion, a textual justification for the score was provided, highlighting reflections on the observed issues. This information was then analyzed to answer \textbf{RQ2}. 
More specifically, for each generated diagram, A1 and A2 included the following information in the evaluation log: (1) the change applied to the requirements, if any; (2) the evaluation score according to the above criteria, and the textual justification for the scores; (3) additional notes on the observed issues; 
%(4) possible tentative solutions to be better explored in the following phase; 
(4) a link to the conversation with ChatGPT or its textual copy. We make the logs and other related material available in our online annex~\cite{replicationpkg}. 
%alternatively the PlantText code corresponding to the automatically generated diagram. 
%All this information was stored in a textual file for each generated diagram.   
%

In our evaluation, we did not use a manually defined ground truth for two reasons: (a) more than one diagram exists that satisfies the same requirements; (b) existing ground truths are limited (e.g., three diagrams in~\cite{jahan2021generating}), and here we wanted to have a wider perspective on possible issues. Additional reflections on the rationale of the study design are in Sect.~\ref{sec:discussion}.

%% file: sections/tab-collection.tex
\begin{table}[t]
\caption{Data Collection Results}\label{tab:data-col}
\begin{threeparttable}[t]
  \centering
  \begin{tabularx}{0.48\textwidth}{@{} p{0.15\textwidth} @{\hskip 0.5em} p{0.17\textwidth} @{\hskip 0.5em} *{3}{>{\arraybackslash}X}@{}}
  \toprule
\textbf{File}\tnote{*} & \textbf{Domain} & \textbf{REQ}\tnote{$\dag$} & \textbf{VAR}\tnote{$\dag$} & \textbf{ANN}\tnote{$\dag$} \\ 
\midrule
% Triplex (s) & Cyber-physical System  &  8 & 13 & Both  \\ 
% Inventory (s) & Inventory System & 22 & 3 & A2  \\ 
% Autopilot (s) & Cyber-physical System  & 14 & 9 & Both \\ 
% qheadache (s) & Gaming  & 11 & 5 & Both \\
% CentralTradingSys (uc) & E-commerce & 1(5)\tnote{$\ddag$} & 1 & A2 \\
% wrac III (s) &  Archiving &  6 & 3 & A2\\
% datahub (us) & Data Management & 67 & 3 & A2 \\
% g02-uc-cm-req (uc) & Healthcare & 1(11) & 1 & A2\\
% g04-uc-req (uc) & Traffic Control & 1(8) & 3 & A2\\
% g05-uc-req (uc) & Football Digital System & 5(37) & 2 & A2 \\
% pacemaker (s) & Healthcare & 289 & 2 & A2 \\
% UHOPE (us) & Healthcare & 12 & 5 & A2 \\
% FiniteStateMachine (s) & Cyber-physical System  & 13 & 1 & A1 \\
% TustinIntegrator (s) & Cyber-physical System  & 4 & 1 & A1\\
% Regulators (s) & Cyber-physical System  & 10 & 1 & A1\\
% NonlinearGuidance (s) & Cyber-physical System  &  7 & 1 & A1\\
% NeuralNetwork (s) & Cyber-physical System  & 4 & 1 & A1 \\
% EffectorBlender (s) & Cyber-physical System  & 5 & 1 & A1\\
% Euler (s) & Cyber-physical System  & 8 & 1 & A1 \\
% caiso (s) &  Black Start Generation & 6 & 2 & A1\\
% eirene (s) & Railway & 8 & 3 & A1\\
% ertms (s) & Railway & 6  & 6 & A1 \\
% evla-back (s) & Astronomy & 8 & 1 & A1 \\
% g04-recycling (us) & Recycling System & 51 & 3 & A1\\
% g12-camperplus (us) & Camping System & 13 & 2 & A1 \\
% keepass (uc) & Security & 1(11) & 3 & A1 \\
% peering (uc) & Networking & 1(5) & 2 & A1\\
% pnnl (uc) & Energy Diagnostics & 1(11) & 5 & A1 \\
Autopilot (s) & Cyber-physical System  &14&9& Both \\ 
caiso (s) &  Black Start Generation &6&2& A1\\
CentralTradingSys (uc) & E-commerce & 1(5)\tnote{$\ddag$} &1& A2 \\
datahub (us) & Data Management &67&3& A2 \\
EffectorBlender (s) & Cyber-physical System  &5&1& A1\\
eirene (s) & Railway &8&3& A1\\
ertms (s) & Railway &6&6& A1 \\
Euler (s) & Cyber-physical System  &8&1& A1 \\
evla-back (s) & Astronomy &8&1& A1 \\
FiniteStateMachine (s) & Cyber-physical System  &13&1& A1 \\
g02-uc-cm-req (uc) & Healthcare & 1(11)\tnote{$\ddag$} &1& A2\\
g04-recycling (us) & Recycling System &51&3& A1\\
g04-uc-req (uc) & Traffic Control & 1(8)\tnote{$\ddag$} &3& A2\\
g05-uc-req (uc) & Football Digital System & 5(37)\tnote{$\ddag$} &2& A2 \\
g12-camperplus (us) & Camping System &13&2& A1 \\
Inventory (s) & Inventory System &22&3& A2  \\ 
keepass (uc) & Security & 1(11)\tnote{$\ddag$} &3& A1 \\
NeuralNetwork (s) & Cyber-physical System  &4&1& A1 \\
NonlinearGuidance (s) & Cyber-physical System  &7&1& A1\\
pacemaker (s) & Healthcare &289&2& A2 \\
peering (uc) & Networking & 1(5)\tnote{$\ddag$} &2& A1\\
pnnl (uc) & Energy Diagnostics & 1(11)\tnote{$\ddag$} &5& A1 \\
qheadache (s) & Gaming  &11&5& Both \\
Regulators (s) & Cyber-physical System  &10&1& A1\\
Triplex (s) & Cyber-physical System  &8&13& Both  \\ 
TustinIntegrator (s) & Cyber-physical System  &4&1& A1\\
UHOPE (us) & Healthcare &12&5& A2 \\
wrac III (s) &  Archiving &6&3& A2\\
\bottomrule
\end{tabularx}
\begin{tablenotes}
    \item[*] s: ``shall'' requirements; uc: use case specifications; us: user stories.
    \item[$\dag$]  REQ: the number of analyzed requirements, VAR: the number of generated variants, ANN: the annotator who did the analysis.
    \item[$\ddag$] Use Case (Steps): Note that we provide the number of use case specifications considered in the analysis as well as the total number of steps (between parentheses).   
 \end{tablenotes}
 \end{threeparttable}
\end{table}

\iffalse
\begin{table}[t]
\caption{Data Collection Results}\label{tab:data-col}
  \centering
  \begin{tabularx}{0.48\textwidth}{@{} p{0.11\textwidth} @{\hskip 0.5em} p{0.07\textwidth} @{\hskip 0.5em} p{0.09\textwidth} @{\hskip 0.5em} 
  p{0.09\textwidth} @{\hskip 0.5em} *{3}{>{\arraybackslash}X}@{}}
  \toprule
\textbf{Type} & \textbf{Documents} & \textbf{Variants} \\ 
\midrule
``Shall''-style & 21       & 65           \\ 
\midrule
User stories & 4 & 8            \\ 
\midrule
Use cases & 3 & 9\\ 
\bottomrule
\end{tabularx}
\vspace{-2em}
\end{table}
\fi

%% file: sections/results.tex
\section{Execution and Results}\label{sec:results}

\subsection{RQ1: Quality of the Diagrams}
\begin{figure} [t!]
\centering
\includegraphics[width=0.48\textwidth]{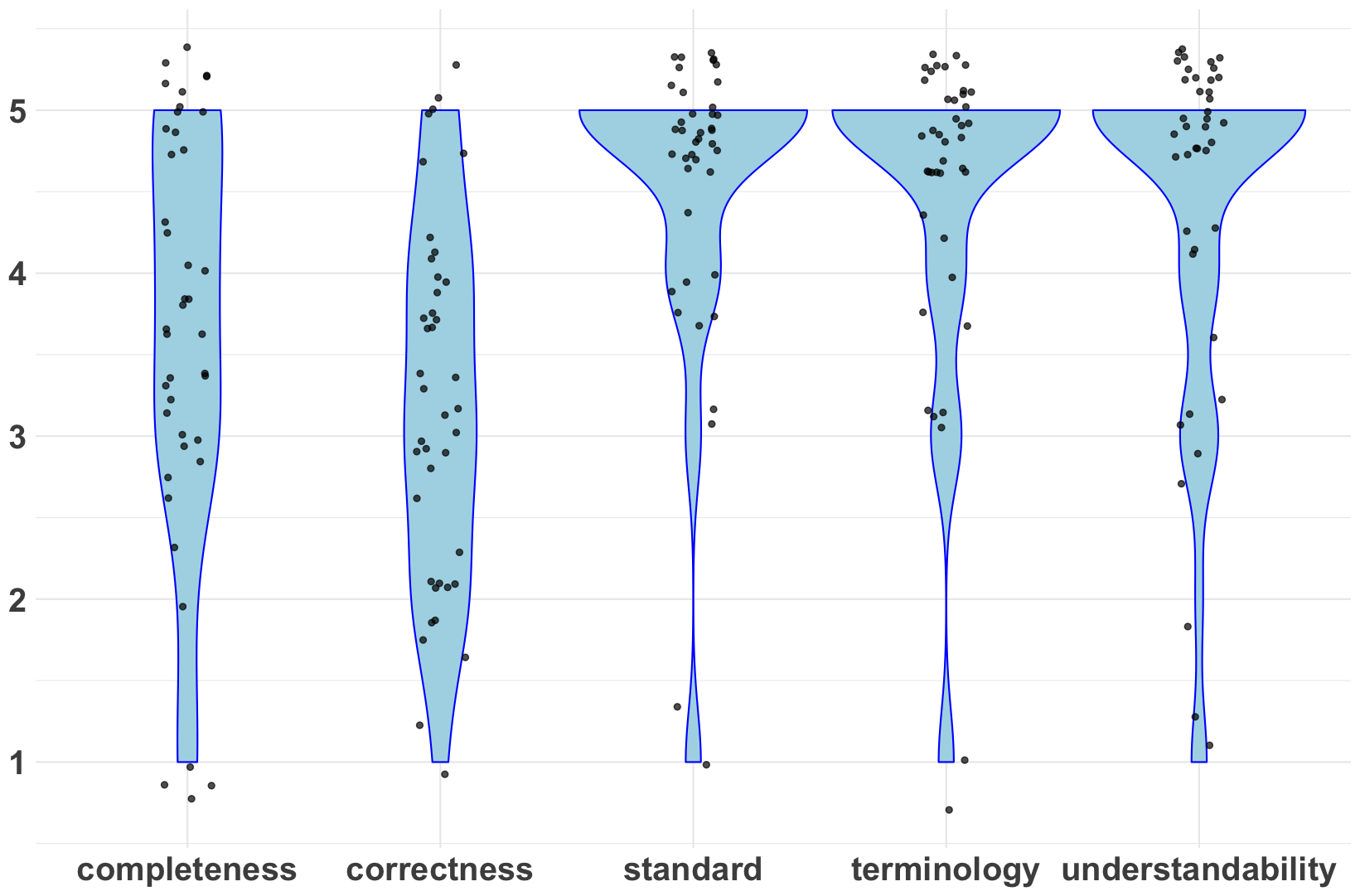}
\caption{Violin plots for the different evaluation criteria.}
\label{fig:boxplots}
\vspace*{-0.5em}
\end{figure}

%\subsection{RQ1: Reliability of ChatGPT}
Fig.~\ref{fig:boxplots} shows the violin plots with jittered points resulting from evaluating the different criteria on the generated models, considering well-formed requirements as input. 
Table~\ref{tab:statistics} reports the statistics and the results of the tests. We see that completeness, understandability, adherence to the standard, and terminological alignment are significantly above the mean value (p-value $\le 0.05$) with medium (d $>0.2$) to large (d~$>0.8$) effect size---reference values from Cohen~\cite{cohen2013statistical}. This indicates a sufficient degree of model quality for these criteria. Instead, scores for correctness issues do not significantly differ from the mean (p-value $> 0.05$), meaning that, on average, the criterion was not adequately fulfilled. In addition, by looking at the violin plot, the distribution of completeness, although skewed towards high values, is still not optimal, with several 3 and 4 scores. 
%A1 and A2 were undecided on the fulfillment of the criterion. 
These results suggest that a requirements analyst who wants to generate diagrams with ChatGPT needs to be careful and check its output's correctness and completeness, paying attention to the issues we outline in the following section.
%, and we fail to reject the hypothesis \textit{The scores for correctness do not differ from the mean value}. 

\input{sections/tab-stats}

%The median values are 4 (completeness), 3 (correctness), and 5 (understandability, adherence to the standard, and terminological alignment), suggesting a sufficient level of reliability for most of the criteria.  

\subsection{RQ2: Emerging Issues}
\label{sec:rq2}
%\subsection{Emerging Issues when Using ChatGPT}
%\hl{Alessio: at the moment the initial paragraph provides explanation of why we do not go in-depth into differences, so that only a qualitative analysis is justified.}
%Fig.~\ref{fig:themes} summarizes the main themes that emerged from the thematic analysis (HL and ML codes only, see replication package~\cite{replicationpkg} for LL codes).
%The categories are the ones used for the evaluation, plus an additional one called \textit{General Issues}, including problems that cannot be directly associated with the evaluation categories.  
%Low-level codes are reported in our \hl{codebook}~\cite{}. 
In the following, we discuss HL and ML codes emerging from the thematic analysis (see~\cite{replicationpkg} for LL codes), also presenting evidence in the form of requirements and diagram excerpts. Please note that each example is a \textit{portion} of a requirements group and a \textit{zoom} of the associated diagram. Complete artifacts are in our evaluation logs~\cite{replicationpkg}.
%, which cannot reported in their entirety for space reasons (see~\cite{replicationpkg} for complete diagrams and requirements). 

%We do not differentiate among different requirements formats. Furthermore, we do not trace requirements modifications with issues. The analysis of format-related issues and the correlation between issues and requirements modifications requires a more systematic exploration and analysis, which is out of the scope of this study. Here, we are mainly interested in general common issues that can inspire research questions for future investigations. 

%new research questions that can be statistically tested in subsequent studies.  
%\hl{Ale: Chetan is right when he says that one wants to know to what extent good models can be generated. So, let us still consider the option of cross-checking only the documents without mutants, and presenting the numerical results for them.}
%We next discuss the main themes from Fig.~\ref{fig:themes}.

\sectopic{1. Completeness.} 
%The generated models appear to show a sufficient degree of completeness with respect to the requirements, although relevant issues are observed concerning missing information. 

\begin{figure} [t!]
\centering
\vspace{-0.5em}
\includegraphics[width=0.4\textwidth]{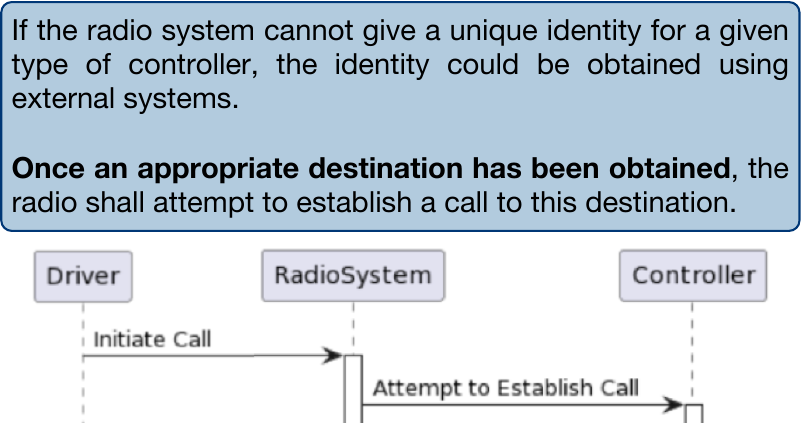}
\caption{``Summarization Issues''.}
\label{fig:summarizationissues}
\vspace{-1.5em}
\end{figure}

\textit{Summarization Issues:} Observations highlight instances where the generated content tends to be excessively summarized.
%, potentially leading to loss of critical information. 
Missing elements encompass requirements or fragments thereof, components, function calls, conditions, and messages. While sequence diagrams unavoidably summarize the content of the requirements, as they are meant to complement and not replace them, it is important to ensure that all the \textit{relevant} information is present. It appears that ChatGPT finds it hard to distinguish between relevant and inconsequential information. 
An example is represented in Fig.~\ref{fig:summarizationissues}, concerning a radio system for a train. We see that the first requirement and the statement ``Once an appropriate destination has been obtained'' are ignored, and the Driver initiates the call, assuming that the destination (i.e., the Controller's identity) was already obtained. 

%\begin{figure} [t!]
%\centering
%\includegraphics[width=0.5\textwidth]{img/ThematicAnalysis.pdf}
%\caption{Themes emerging from the analysis.}
%\label{fig:themes}
%\vspace{-1.5em}
%\end{figure}

%concerning a requirement document for a failure management system with triple redundancy. We see that the condition ``If a second failure is in progress'' is ignored, and the selected value is maintained regardless of the failure state. 

%0 Triplex\_v1.txt

%\hl{to be re-coded as incorrect structure}

%\textit{Missing Abstractions:} Besides missing elements that appear in the requirements, ChatGPT is sometimes not sufficiently effective in capturing conceptual abstract elements, such as states and control loops. Sometimes, these abstractions are not explicitly stated in the requirements, but one would expect an expert modeler to introduce them in a sequence diagram that provides a conceptual perspective of the system behavior. 
%An example is Fig.~\ref{fig:missingabstractions}. Besides missing an explicitly mentioned state (no-fail state), the diagram misses a control loop, which should be introduced as all the required actions are performed cyclically. The model also includes other issues described later.

%0 Triplex\_v5.txt

\begin{figure} [!h]
\centering
\vspace{-1em}
\includegraphics[width=0.4\textwidth]{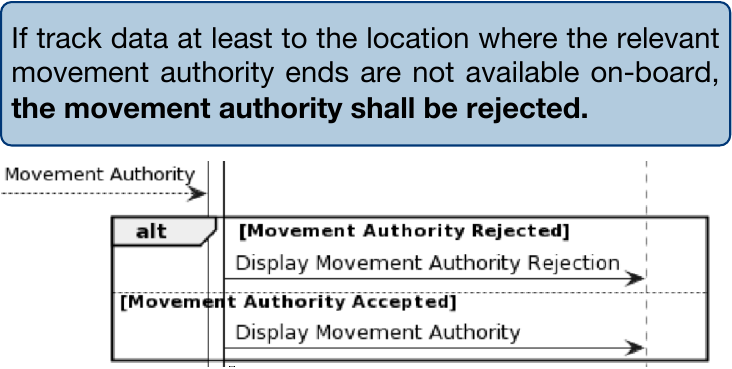}
\vspace{-0.5em}
\caption{``Poor Requirements Quality and Model Omissions''.}
\label{fig:reqqualityomissions}
\vspace{-0.25em}
\end{figure}

\textit{Poor Requirements Quality and Model Omissions:} One interesting phenomenon is the possible effect of requirements quality on the omission of relevant information. It appears that when requirements are hard to understand or ambiguous, the information associated with them is somewhat ``hidden'' by abstracting their details, which leads to incomplete diagrams. In Fig.~\ref{fig:reqqualityomissions}, concerning the 
requirements for a train control system, we see that a condition that is hardly understandable (``If track data at least to the location where the relevant movement authority ends are not available on-board'') is basically concealed behind the alternative ``Movement Authority Rejected''.

%ertms\v03

\begin{figure} [t!]
\centering
\includegraphics[width=0.4\textwidth]{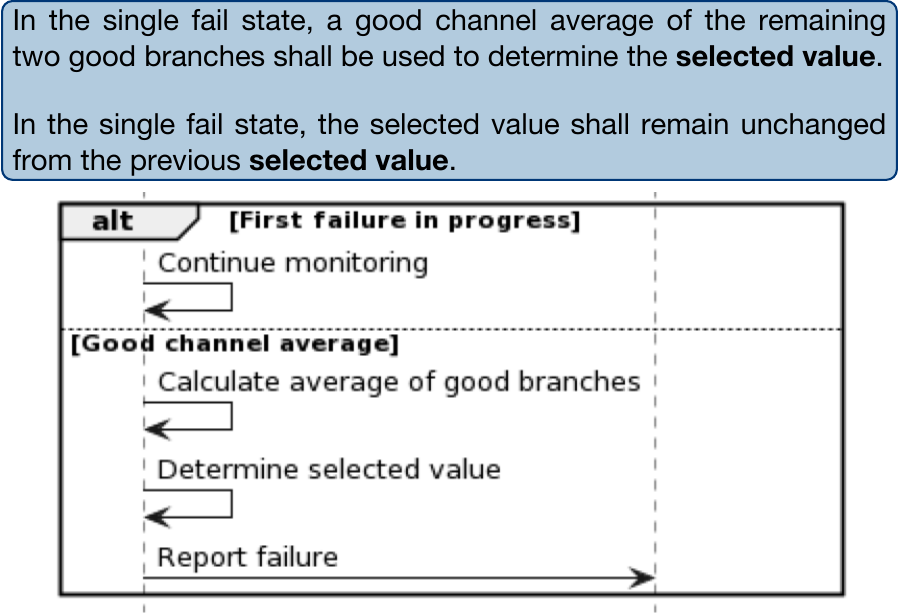}
\caption{``Inconsistency and Model Omissions'', ``Inconsistency and Model Incorrectness'', and ``Incoherence Manifestations''.}
\label{fig:inconsistencyomissions}
\vspace{-0.5em}
\end{figure}

\textit{Inconsistency and Model Omissions:} Similar to the previous code, inconsistency between requirements can be associated with situations in which ChatGPT hides the conflict, thus leading to omissions. Fig.~\ref{fig:inconsistencyomissions} considers a failure management system with triple redundancy. We have two conflicting requirements---different selected values for the same state. The conflict is hidden by the function call ``Determine selected value'', which does not specify how the value should be determined. 

\begin{figure} [!h]
\centering
\vspace{-0.5em}
\includegraphics[width=0.40\textwidth]{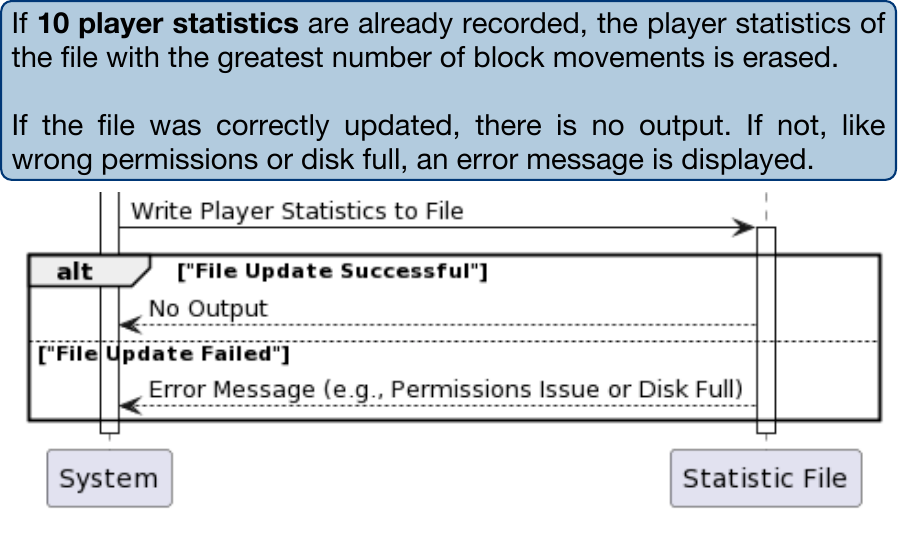}
\vspace{-1.5em}
\caption{``Poor Precision in Timing and Numbers''.}
\label{fig:timingnumbers}
%\vspace{-1em}
\end{figure}

\textit{Poor Precision in Timing and Numbers:} 
As LLMs primarily focus on language, ChatGPT notably encounters difficulties with numbers, timing, and mathematical thinking in general. This challenge becomes evident when the requirements involve numerical constraints, as in Fig.~\ref{fig:timingnumbers}, where the requirements of a computer game are considered. The requirement with the numerical condition (``If 10 player statistics are already recorded'') is entirely ignored.

%qheadachev\_02

\sectopic{2. Correctness.} %With some exceptions, the generated models correctly represent the requirements in most of the cases. However, several correctness issues emerge when requirements have defects, such as incompleteness, inconsistency, and lack of clarity.

%\hl{I will add more cases of incorrectness, as they are too few compared with the scores. Additional issues are I found are: 1. incorrect structure, including action grouping (we can use missing abstractions), 2. incorrect component/participant (caiso from Sallam), 3. incorrect interaction (also order of actions, wrong message flow), 4. complexity leads to incorrectness (nonlinear guidance).}

\begin{figure} [t!]
\centering
\vspace{-0.5em}
\includegraphics[width=0.4\textwidth]{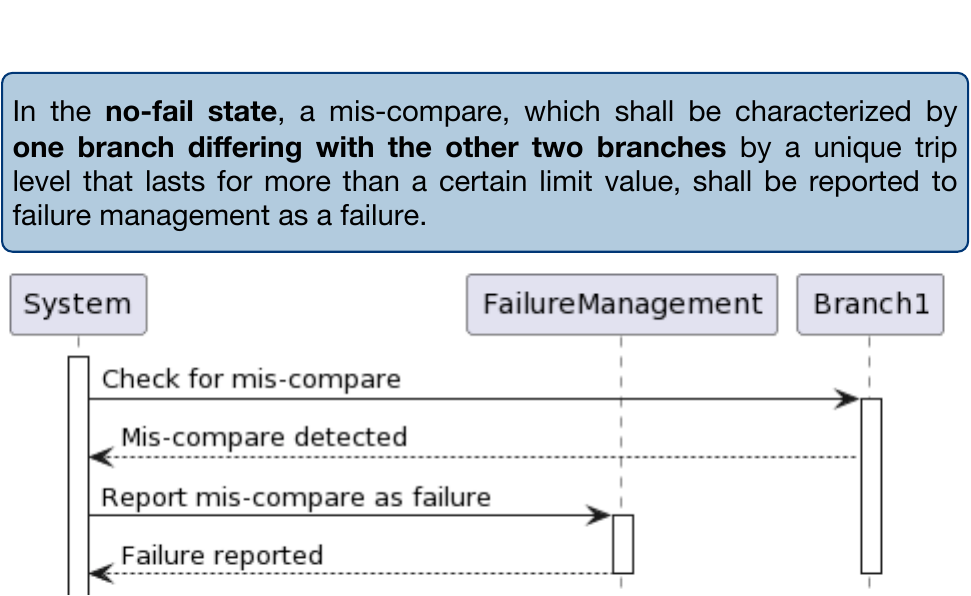}
\caption{``Incorrect Structure'' and ``Incorrect Interaction''.}
\label{fig:missingabstractions}
%\vspace{-1.5em}
\end{figure}

\textit{Incorrect Interactions:} The models can include an incorrect order of function calls/messages or an incorrect flow of actions with respect to what is expressed in the requirement.  
%The sequence diagrams are expected to adhere to the original requirements not just in terms of coverage but also in terms of their correct interpretation and representation. In some cases, the generated output showed behavior that was not specified in the requirements, and that was therefore judged as incorrect. 
This is the case of Fig.~\ref{fig:missingabstractions}, again considering a failure management system with triple redundancy. Here, only one branch is considered to detect a mis-compare, instead of three branches, as the requirement indicates (``one branch differing from the other two branches''). 

%0 Triplex\_v5.txt 

\begin{figure} [t!]
\centering
\includegraphics[width=0.3\textwidth]{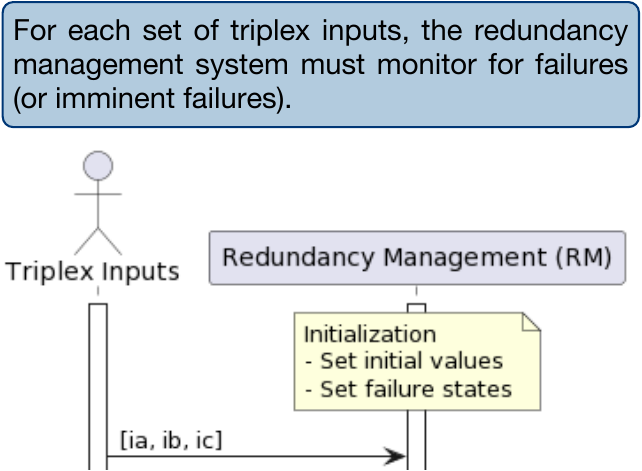}
\caption{``Incorrect Component/Actor''.}
\label{fig:incorrectcomponent}
%\vspace{-0.5em}
\end{figure}

\textit{Incorrect Component/Actor:} The models include components or actors that should be messages or functions. Fig.~\ref{fig:incorrectcomponent} exemplifies this case, where ``Triplex input'' is treated as an actor, while it should be a data structure passed through a message from the sensors. 

%\textbf{Example:} pnnl v4

\textit{Incorrect Structure:}
ChatGPT is sometimes not sufficiently effective in capturing conceptual abstract elements, such as states and control loops, which help to correctly structure the models. In other cases, function calls are not appropriately grouped and are scattered throughout the diagram. 
%, which affects the structure.   
%Sometimes, these abstractions are not explicitly stated in the requirements, but one would expect an expert modeler to introduce them in a sequence diagram that provides a conceptual perspective of the system behavior. 
In Fig.~\ref{fig:missingabstractions}, the ``no-fail state'' is incorrectly overlooked (this can also be regarded as a summarization issue), and the diagram misses a control loop, which is needed as all the actions are performed cyclically. 
%The model also includes other issues described later.

\begin{figure} [!h]
\centering
\vspace{-0.8em}
\includegraphics[width=0.5\textwidth]{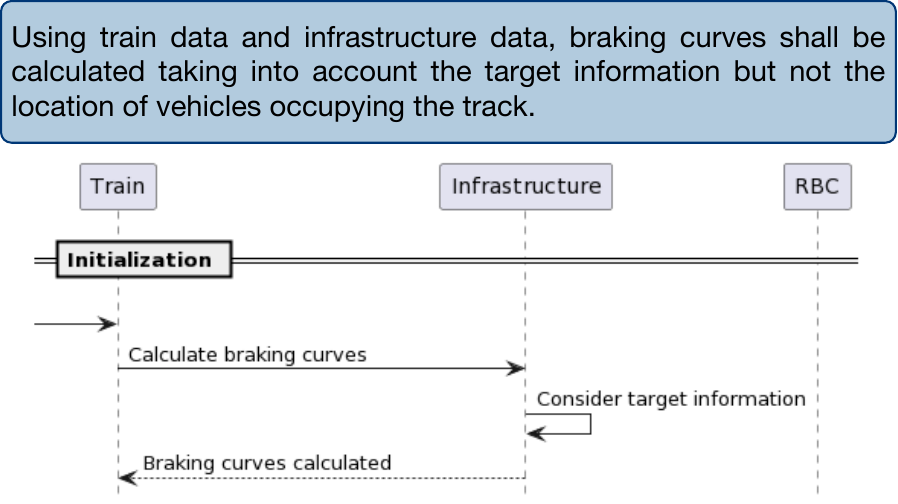}
\vspace{-0.5em}
\caption{``Unclear Requirements and Model Incorrectness''.}
\label{fig:unclearincorrectness}
\vspace{-0.5em}
\end{figure}

\textit{Unclear Requirements and Model Incorrectness:} %Observations highlight the impact of unclear requirements on the accuracy of the generated sequence diagrams. 
When ChatGPT encounters vague or ambiguous aspects in the input requirements, it may produce content open to interpretation, potentially leading to inaccuracies. 
The requirement for a train control system in Fig.~\ref{fig:unclearincorrectness} uses passive voice (``braking curves shall be calculated''), which leads ChatGPT to think that the infrastructure calculates the braking curve of the train, although the train system normally computes it. 

%\textit{Incompleteness Leads to Incorrectness:}
%An observation emerging from the analysis is the possible correlation between the incompleteness of the requirements and the potential introduction of inaccuracies in the generated sequence diagrams. While requirements may include tacit knowledge that can be inferred by a domain expert, omission of information in the input requirements appears to lead to incorrect interpretation and, thus, representation. 

%\textbf{Example}
%\hl{CANNOT FIND a CLEAR ONE, remove?}

\textit{Inconsistency and Model Incorrectness:} 
ChatGPT may struggle to produce accurate representations when encountering inconsistencies or conflicting information. In Fig.~\ref{fig:inconsistencyomissions}, the inconsistency between the requirements is not only hidden but appears to produce incorrect behaviour. When the diagram condition ``First failure in progress'' is true (i.e., ``in the single fail state'', in the requirements' terminology), only the second requirement is considered (i.e., ``the value shall remain unchanged''), while the first is not.

%\textit{Complexity Leads to Incorrectness:} When several requirements enter into play, ChatGPT struggles to make sense of them and creates diagrams that do not adhere to the requirements.  This suggests that there is a limit in terms of the complexity of the input data that the LLM can successfully process. We do not report examples in this case    

%caiso v4
%ertms v05 passive voice

\sectopic{3. Adherence to the Standard.}

%\textit{Use of a Language subset:} Observations suggest that the generated content tends to use simplified language, potentially indicating challenges in adhering to more refined constructs as per standard practices.

\textit{Syntax Errors:} In some cases, the output cannot be interpreted by PlanText, raising syntactic errors. Even after pinpointing the specific error, ChatGPT cannot recover from it, forcing the user to modify the diagram manually. 

\begin{figure} [!h]
\centering
\includegraphics[width=0.3\textwidth]{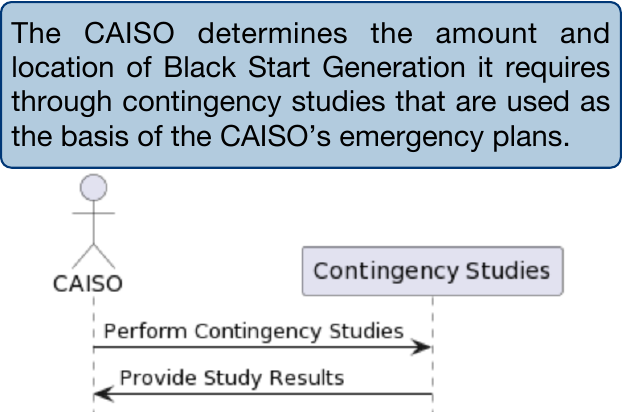}
\caption{``Semantic Errors'' and ``Additional Terms''.}
\label{fig:semantic}
\vspace{-1em}
\end{figure}

\textit{Semantic Errors:} The usage of constructs is sometimes semantically incorrect. A case is the usage of function calls instead of messages, as shown by the call to ``Provide study results'' in Fig.~\ref{fig:semantic}, concerning a system that should have been a return message.

\sectopic{4. Terminology.} 
%Observations reveal that the terminology in the generated content aligns appropriately with the language used in the original requirements. 

\textit{Additional Terms:} Novel terms that were not originally present in the requirements are introduced by ChatGPT, which hinders the coherence of the models with respect to the requirements. This is shown in Fig.~\ref{fig:semantic}, about a standard for energy grids, where ``Study Results'' are never mentioned. 

\textit{Relevant Terms Missing:} Terms that were considered relevant during the analysis are not reported in the model. Again, Fig.~\ref{fig:semantic} shows that the term ``Black Start Generation'', which appears to be relevant for the requirement, as expressed in capital letters, is missing from the diagram.

%\textit{Unclear Terminology:} The meaning of the newly introduced terminology is not sufficiently clear. 

\textit{Inconsistent Terminology:} The terms introduced are not consistent with the original ones, although they appear to express a similar meaning. For example, in Fig.~\ref{fig:inconsistencyomissions}, the expression ``In the single fail state'' appears to be replaced by the condition ``First failure in progress''. It should be, however, acknowledged that terminological issues are often acceptable if they preserve the intended meaning of the original terms.

\sectopic{5. Understandability.} Due to the complexity of models with understandability issues, here we report only codes that can be represented synthetically through diagram excerpts.

\begin{figure} [t!]
\centering
\includegraphics[width=0.3\textwidth]{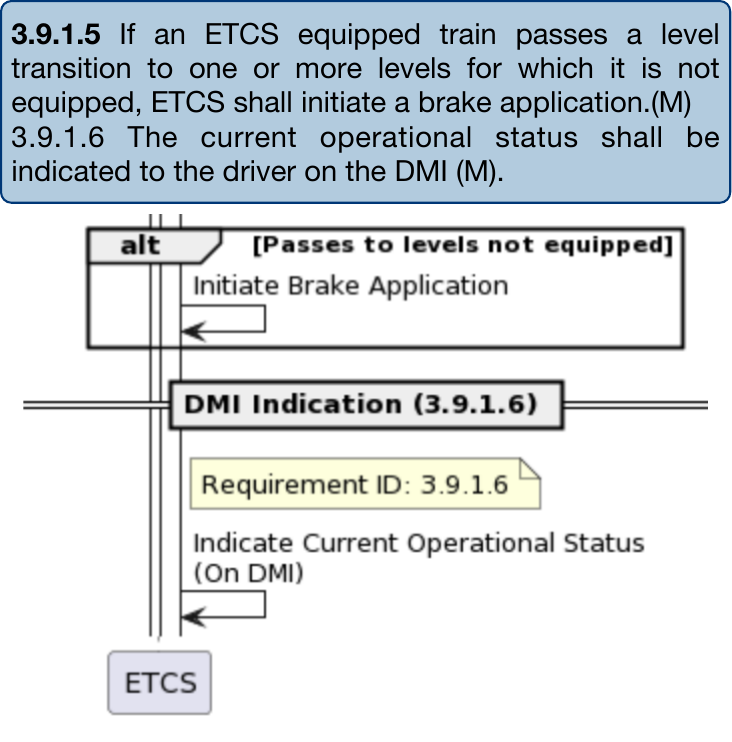}
\caption{``Traceability Challenges''.}
\label{fig:traceability}
%\vspace{-2em}
\end{figure}

\textit{Traceability Challenges:} Assessing completeness and correctness can be hard without clearly tracing the requirements and the diagram. Tracing information is sometimes included as notes, which, although helpful in principle, can be incomplete or inaccurate. Fig.~\ref{fig:traceability} shows a case in which only one requirement is traced while the other is ignored, although the associated action is correctly displayed (``Initiate Brake Application'').

\textit{Incoherence Manifestations:} When requirements are inconsistent, unclear, or incomplete, ChatGPT attempts to produce a model, but often at the cost of understandability, e.g., in Fig~\ref{fig:inconsistencyomissions}, it is unclear what ``Good channel average'' means. 

\begin{figure} [t!]
\centering
\includegraphics[width=0.44\textwidth]{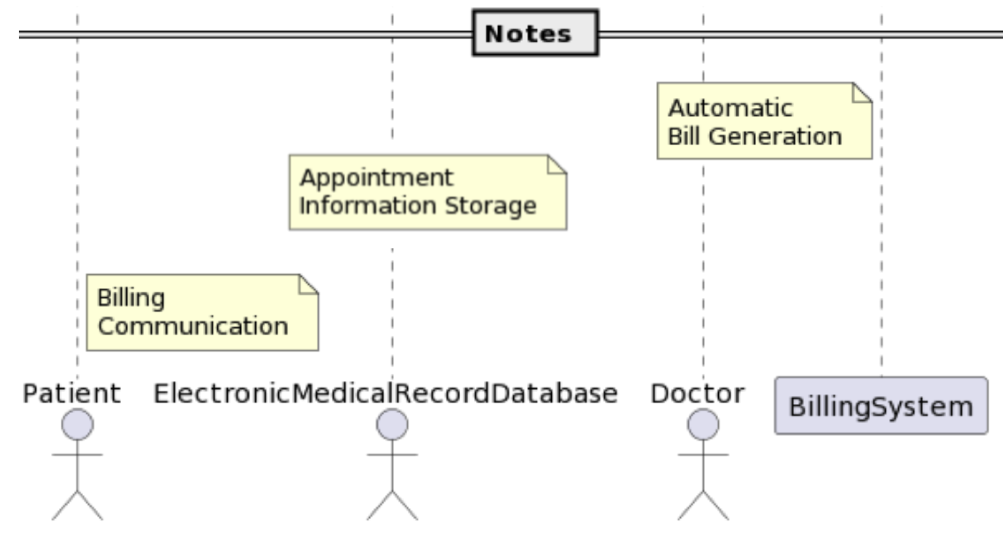}
\vspace{-.5em}
\caption{``Misplaced Information''.}
\label{fig:misplaced}
\vspace{-1.5em}
\end{figure}

\textit{Misplaced Information:} Sometimes, clarifying notes are added, which can be useful to facilitate understandability (cf. Fig.~\ref{fig:incorrectcomponent}). However, they are occasionally not placed close to the model parts they refer to, as in Fig.~\ref{fig:misplaced}, where notes are aggregated in the bottom part of the diagram. 

\textit{Presence of Redundancy:} ChatGPT can introduce redundant or superfluous information, which, although it does not impact the correctness of the model, affects its understandability. 

%\textit{Superfluous Information:} Observations indicate the presence of unnecessary concepts, which affect the overall understandability.

\textit{Actor Overrepresentation:} Sometimes too many actors are introduced, which makes the sequence diagram too wide and hard to navigate. 

\textit{Too Detailed:} The diagrams should faithfully represent the requirements, but sometimes the generated diagrams are too detailed with nested conditions and loops to be easily navigated. In such cases, a higher degree of abstraction would be expected. 

%\textit{Absence of Rationale:} Noteworthy is the absence of rationale within the generated content, suggesting potential gaps in providing context and reasoning. This occurs especially in user stories.

%\sectopic{Key Issues Stemming from LLMs Usage.}

\sectopic{6. General Issues.}

%\textit{Reliance on Pre-trained Information of Sequence Diagrams:}
%One of the key issues associated with sequence diagrams (or, for that matter, any diagram generation) using LLMs is the reliance on the pre-trained information of `what is a sequence diagram?'. We noticed several issues of completeness, correctness, and adherence to standard are associated with ChatGPT occasionally faltering at the precise understanding of sequence diagrams. For instance, in a few cases, the generated sequence diagrams had syntax issues and incorrect usage of constructs (as discussed in adherence to the standard). 

\textit{Memory-Induced Hallucinations:} 
Occasionally, ChatGPT appears to hallucinate, i.e., it provides an output inconsistent with the query. We observed that this is likely due to interactions with the user. Specifically, we sometimes observed that requirements used in previous independent sessions and generated models appeared to influence subsequent sessions. 
%For example, when a class diagram was generated in a previous session, ChatGPT generated this type of diagram even if prompted to produce a sequence diagram. 

\textit{Ignored Requirements Modifications:} Possibly due to ChatGPT's memory of previous chat sessions, modifications to the requirements are sometimes ignored. This can complicate an %empirical evaluation such as ours and an 
iterative process in which requirements are incrementally introduced, and models are generated through multiple iterations, as in the scenario illustrated by~\cite{ambriola2006systematic}.  

\textit{Lack of Contextual Understanding:} ChatGPT might not have domain-specific knowledge or context, which can be crucial for accurately translating requirements into sequence diagrams. Given the token limit of ChatGPT, it might not be easy to provide the context required for ChatGPT to represent a sequence diagram accurately. For instance, in some cases where requirements had cross-references to other parts, the generated output exhibited lower quality, possibly due to the lack of context. 
Another case is Fig.~\ref{fig:missingabstractions}, where ChatGPT appears to miss the technical knowledge of how a triple redundancy failure management system should normally work. 

\textit{Variability of the Output:} Given the same query and a set of requirements, the output of ChatGPT can largely vary. Although this is a well-known issue~\cite{camara2023assessment}, 
%which affects the overall reliability of the instrument. However, 
it should be noted that different diagram representations can be appropriate for the same requirements, and the generation of alternatives can even allow the user to choose a preferred one. Variability in multiple runs is not a substantial problem \textit{per-se}, but rather an asset, if the correctness of the output is ensured. 

%% file: sections/tab-stats.tex
\iffalse
\begin{table}[t]
\caption{Statistics and test results}\label{tab:statistics}
\centering
  \footnotesize
  \begin{tabularx}{0.48\textwidth}{@{} p{0.11\textwidth} @{\hskip 0.5em} p{0.07\textwidth} @{\hskip 0.5em} p{0.09\textwidth} @{\hskip 0.5em} 
  p{0.09\textwidth} @{\hskip 0.5em} *{1}{>{\arraybackslash}X}@{}}
  \toprule
\textbf{Criterion} & \textbf{Mean} & \textbf{p-value}  & \textbf{Cohen d} & \textbf{Effect Size} \\ \midrule
completeness       & 3.5854        & \textbf{0.0088}   & 0.4248           & Medium             \\ 
\midrule
correctness        & 3.1707        & 0.1282            & 0.1374           & Small              \\ \midrule
standard           & 4.4878        & \textbf{1.88E-06} & 1.2776           & Large              \\ \midrule
understand. & 4.3171        & \textbf{8.32E-06} & 1.0342           & Large              \\ 
\midrule
terminology        & 4.4390        & \textbf{3.52E-06} & 1.1940           & Large              \\ 
\bottomrule
\end{tabularx}
\vspace{-1em}
\end{table}
\fi

\begin{table}[t]
\caption{Statistics and test results}\label{tab:statistics}
\centering
\begin{tabular}{|l|r|r|r|l|} 
\hline
\multicolumn{1}{|c|}{\textbf{Criterion}} & \multicolumn{1}{c|}{\textbf{Mean}} & \multicolumn{1}{c|}{\textbf{p-value}} & \multicolumn{1}{c|}{\textbf{Cohen d}} & \multicolumn{1}{c|}{\textbf{Eff. S.}}  \\ 
\hline
complet.                             & 3.634146                           & 0.003652                              & 0.503258                              & Med.                                     \\ 
\hline
correctness                              & 3.219512                           & 0.101543                              & 0.194357                              & Small                                      \\ 
\hline
standard                                 & 4.536585                           & 1.4E-07                               & 1.572471                              & Large                                      \\ 
\hline
understand.                        & 4.365854                           & 6.35E-07                              & 1.22766                               & Large                                      \\ 
\hline
terminology                              & 4.487805                           & 2.13E-07                              & 1.447751                              & Large                                      \\
\hline
\end{tabular}
\end{table}

%% file: sections/discussion.tex
%\vspace*{-.8em}
\section{Discussion}
\label{sec:discussion}

In the following, we discuss our findings, %comparing them with related works, and 
and speculate on possible solutions to the identified problems. 

\textit{Requirements Quality and Diagram Quality.} Correctness of the diagrams is the main issue observed, followed by completeness. We noticed that these aspects could be associated with requirements quality issues, as ChatGPT could neglect or conceal requirements content that cannot be interpreted unequivocally. To address this issue, one should consider performing quality checks on the requirements, either through manual inspection,
%~\cite{shull2000perspective}
automatic tools~\cite{zhao2021natural}, or through the support of ChatGPT itself, as done in recent works~\cite{Arora:23}. 
%,zhang2023preliminary}. 
The relation between requirements defects and diagram quality should be confirmed by more rigorous experiments stemming from our exploratory study. If this relation is confirmed, diagram generation with ChatGPT could help spot requirements quality issues. 
%, as poor requirements could be associated with low-quality diagrams. 
ChatGPT could play the role of a fictional conversational partner that strives to comprehend the content of the requirements and then demonstrates its understanding through the creation of diagrams. 
%These are expected to exhibƒprobinhΩΩit higher quality when the requirements are clearly interpretable, otherwise, 
The generation of a low-quality diagram may suggest that further improvement of the requirements is needed. 

%The potential of ChatGPT in the different requirements tasks can be exploited to create an LLM-powered RE process, as suggested... 

\textit{Incremental/Interactive Diagram Definition.}
An RE process in which ChatGPT acts as an assistant does not rule out the primary role of an experienced requirements analyst. Multiple prompt iterations can be needed before ChatGPT can produce a diagram in the desired form and with the expected quality. These iterations should help ChatGPT self-correct its output and address more complex real-world issues. In practical contexts, requirements may need to be first decomposed by functionality and then transformed into clear lists of steps amenable to a sequential representation. 
%Investigating prompts for requirements transformation is a possible avenue for research. 
Furthermore, other types of diagrams (e.g., domain models in the form of class diagrams and goal models) may be required to create higher-level abstractions or different views. ChatGPT can provide support in this regard~\cite{Arora:23,ronanki2023investigating,chen2023use,camara2023assessment}. In other contexts, one may start from early requirements and incrementally use ChatGPT to produce diagrams that help to refine the requirements further~\cite{ambriola2006systematic}. Incremental prompt engineering strategies are needed to accommodate these diverse RE contexts and identify the best way to exploit the synergy between requirements analysts and LLMs.   
%Current work on prompt engineering for RE should pursue this avenue  

%also asking ChatGPT to self-correct its behavior. 

\textit{The Role of Domain/Contextual Knowledge.} We have observed that requirements that included term definitions and some additional context allowed ChatGPT to provide richer and more accurate representations. While ChatGPT has been trained on large amounts of documents, it may lack the technical knowledge often required to comprehend the requirements. Within the token limits, partial context and knowledge can be provided in a prompt.  When one uses the ChatGPT API, fine-tuning with context-specific data is also possible. Furthermore, probing prompts can be used first to verify whether the LLM correctly understands a requirement (e.g., asking to rephrase it). When its understanding appears to be incorrect or only partial, the user can provide the additional domain knowledge that the LLM lacks. 
%Tacit knowledge is a well-known problem in RE~\cite{ferrari2015ambiguity}, and interaction with an artificial domain ignorant~\cite{niknafs2017impact} could help make it explicit. 

%In other cases, we have seen that the lack of domain-specific or contextual knowledge could affect the correct interpretation of the requirements, leading to clerical errors (e.g., a Triplex Input considered as an actor in Fig.~\ref{fig:incorrectcomponent}). 
%Tacit knowledge is a well-known problem in RE~\cite{}, and  

\textit{Improving Understandability.} Understandability of diagrams is generally good. However, an apparently clear diagram might lead an analyst to believe it is correct, even if it is not. %This can be addressed by introducing notes and traceability information that clarify the relationship between requirements and models. 
ChatGPT spontaneously introduced notes and traceability information, which---although sometimes imperfect---can greatly help the interpretation of diagrams. 
Explicitly requesting these explanatory elements could provide better transparency, aiding users in comprehending the rationale behind certain decisions or representations, and assessing correctness. ChatGPT is also notoriously verbose in its answers, but having NL explanations---besides diagrams with notes---and explicitly asking for them can provide useful information. This can be used as additional documentation for the diagrams when these are exchanged with stakeholders.  

\textit{Empirical Research on LLMs.} We acknowledge that the research design adopted in this study is unorthodox. However, the study matter is novel, and appropriate validation methods for LLMs are still under development~\cite{sallou2023breaking}. We believe that our work contributes with a research design that can be adopted by other studies using LLMs. Following the empirical research framework by Stol and Fitzgerald~\cite{stol2020guidelines}, our design is positioned between a judgment study (where subject matter experts express their opinions) and a sample study (where objects/subjects are sampled from a population and analyzed/surveyed). These studies aim for generalizability of the findings over different contexts but cannot inherently control the behavior of subjects/objects (as, e.g., in experiments) and cannot achieve context specificity (as, e.g.,  in case studies). We deem this design appropriate for situations in which one cannot fully control a phenomenon, such as ChatGPT's behavior, and wants to perform an exploratory investigation over different realistic situations, i.e., different requirements sample cases.  

%% file: sections/threats.tex
\section{Threats to Validity}
\label{sec:threats}

\textit{Construct Validity.} This study involved a quantitative assessment of ChatGPT's reliability based on five criteria, evaluated using a scoring scale. The use of a scoring scale introduces a potential subjectivity threat in criteria interpretation and evaluation. To address this, we established a shared definition of the criteria, based on a selection of model-relevant criteria inspired by~\cite{8559686}, and preliminary experiments. More criteria can be added in future works. We also computed a Cohen's kappa on a subset of the models, indicating substantial agreement, which mitigates the subjectivity threat.     

\textit{Internal Validity.} A1 and A2 did not use a gold standard, so they used their judgment to assess that the models were actually faithful to the requirements. This subjectivity threat cannot be entirely mitigated, but it is arguably justified by the exploratory nature of the study, and mitigated by the experience of the assessors in RE. Furthermore, the evaluation of A1 and A2 was documented through NL logs, a process inherently subjective. To enhance objectivity, the thematic analysis was conducted by A3, who was not part of the initial evaluation, providing an impartial perspective. To ensure coherence, A1 and A2 contributed to the sanitization and consolidation of themes, forming a triangulation approach that partially mitigates subjectivity threats. The ChatGPT memory of previous sessions (cf. Sect.~\ref{sec:results}) could have also affected the results. However, this is also the behavior that one should expect in practice, and thus makes our evaluation more realistic.

\textit{External Validity.} Our results stem from an exploration of the ability of ChatGPT to generate sequence diagrams from requirements. The exploration is nonsystematic and unavoidably incomplete. The authors performed their analyses independently, which mitigates the threat of an incomplete exploration, as they introduced different variants in the original models, and considered different documents. The selected documents could have biased the evaluation. However, we considered different requirement types in our exploration, which extends the scope of validity of our conclusions, and their generalizability.

%From data collection: 
%the authors performed their explorations independently, which mitigates the threat of an incomplete exploration.

%\cite{wohlin2003empirical}

%% file: sections/conclusion.tex
\section{Conclusion}~\label{sec:conclusion}
In this study, we explored the reliability of ChatGPT in generating UML sequence diagrams from NL requirements. The evaluation revealed promising results in terms of terminological consistency, understandability, and adherence to the standard of the generated models. 
%ChatGPT also demonstrated a sufficient level of reliability in maintaining terminological alignment with the input requirements. 
However, challenges emerged in terms of model \textit{completeness} and \textit{correctness}, particularly when dealing with ambiguous or inconsistent requirements. 
%Syntax and semantic issues were identified, impacting the model's adherence to established standards. Understandability varied, with notable issues such as actor overrepresentation and challenges in traceability. 
We also observed occasional hallucinations, and, in some instances, the model demonstrated limitations in contextual understanding and appeared to lack domain-specific knowledge. 
These findings have important implications for leveraging LLMs in RE. Providing additional contextual information and more domain knowledge can be expected to improve ChatGPT's model generation capability. 
Furthermore, devising incremental prompting strategies with the human-in-the-loop to decompose, verify, and refine the requirements, can help to enhance the correctness of the produced models. 
%Overall, we argue that LLMs can greatly improve and speed up the RE process, but more studies are required to better \textit{control} this powerful machinery. 

%% file: paper.bbl
\begin{thebibliography}{10}
\providecommand{\url}[1]{#1}
\csname url@samestyle\endcsname
\providecommand{\newblock}{\relax}
\providecommand{\bibinfo}[2]{#2}
\providecommand{\BIBentrySTDinterwordspacing}{\spaceskip=0pt\relax}
\providecommand{\BIBentryALTinterwordstretchfactor}{4}
\providecommand{\BIBentryALTinterwordspacing}{\spaceskip=\fontdimen2\font plus
\BIBentryALTinterwordstretchfactor\fontdimen3\font minus
  \fontdimen4\font\relax}
\providecommand{\BIBforeignlanguage}[2]{{%
\expandafter\ifx\csname l@#1\endcsname\relax
\typeout{** WARNING: IEEEtran.bst: No hyphenation pattern has been}%
\typeout{** loaded for the language `#1'. Using the pattern for}%
\typeout{** the default language instead.}%
\else
\language=\csname l@#1\endcsname
\fi
#2}}
\providecommand{\BIBdecl}{\relax}
\BIBdecl

\bibitem{jolak2020software}
R.~Jolak, M.~Savary-Leblanc, M.~Dalibor, A.~Wortmann, R.~Hebig, J.~Vincur,
  I.~Polasek, X.~Le~Pallec, S.~G{\'e}rard, and M.~R. Chaudron, ``Software
  engineering whispers: The effect of textual vs. graphical software design
  descriptions on software design communication,'' \emph{Empirical software
  engineering}, vol.~25, pp. 4427--4471, 2020.

\bibitem{wagner2019status}
S.~Wagner, D.~M. Fern{\'a}ndez, M.~Felderer, A.~Vetr{\`o}, M.~Kalinowski,
  R.~Wieringa, D.~Pfahl, T.~Conte, M.-T. Christiansson, D.~Greer \emph{et~al.},
  ``Status quo in requirements engineering: A theory and a global family of
  surveys,'' \emph{ACM TOSEM}, vol.~28, no.~2, pp. 1--48, 2019.

\bibitem{ambriola2006systematic}
V.~Ambriola and V.~Gervasi, ``On the systematic analysis of natural language
  requirements with {CIRCE},'' \emph{ASE}, vol.~13, pp. 107--167, 2006.

\bibitem{arora2019active}
C.~Arora, M.~Sabetzadeh, S.~Nejati, and L.~Briand, ``An active learning
  approach for improving the accuracy of automated domain model extraction,''
  \emph{ACM TOSEM}, vol.~28, no.~1, pp. 1--34, 2019.

\bibitem{horkoff2019goal}
J.~Horkoff, F.~B. Aydemir, E.~Cardoso, T.~Li, A.~Mat{\'e}, E.~Paja,
  M.~Salnitri, L.~Piras, J.~Mylopoulos, and P.~Giorgini, ``Goal-oriented
  requirements engineering: an extended systematic mapping study,'' \emph{REJ},
  vol.~24, pp. 133--160, 2019.

\bibitem{uml}
``Unified modeling language {({UML})} 2.5.1 core specification,''
  \url{https://www.omg.org/spec/{UML}}, 2017.

\bibitem{yue2015atoucan}
T.~Yue, L.~C. Briand, and Y.~Labiche, ``{aToucan}: an automated framework to
  derive {UML} analysis models from use case models,'' \emph{ACM TOSEM},
  vol.~24, no.~3, pp. 1--52, 2015.

\bibitem{jahan2021generating}
M.~Jahan, Z.~S.~H. Abad, and B.~Far, ``Generating sequence diagram from natural
  language requirements,'' in \emph{REW'21}.\hskip 1em plus 0.5em minus
  0.4em\relax IEEE, 2021, pp. 39--48.

\bibitem{saini2022automated}
R.~Saini, G.~Mussbacher, J.~L. Guo, and J.~Kienzle, ``Automated, interactive,
  and traceable domain modelling empowered by artificial intelligence,''
  \emph{SoSym}, pp. 1--31, 2022.

\bibitem{Arora:23}
C.~Arora, J.~Grundy, and M.~Abdelrazek, ``Advancing requirements engineering
  through generative {AI}: Assessing the role of {LLMs},'' \emph{arXiv preprint
  arXiv:2310.13976}, 2023.

\bibitem{ahmed2022automatic}
S.~Ahmed, A.~Ahmed, and N.~U. Eisty, ``Automatic transformation of natural to
  unified modeling language: A systematic review,'' in \emph{SERA'22}.\hskip
  1em plus 0.5em minus 0.4em\relax IEEE, 2022, pp. 112--119.

\bibitem{chen2023use}
B.~Chen, K.~Chen, S.~Hassani, Y.~Yang, D.~Amyot, L.~Lessard, G.~Mussbacher,
  M.~Sabetzadeh, and D.~Varr{\'o}, ``On the use of {GPT-4} for creating goal
  models: an exploratory study,'' in \emph{REW'23}.\hskip 1em plus 0.5em minus
  0.4em\relax IEEE, 2023, pp. 262--271.

\bibitem{camara2023assessment}
J.~C{\'a}mara, J.~Troya, L.~Burgue{\~n}o, and A.~Vallecillo, ``On the
  assessment of generative {AI} in modeling tasks: an experience report with
  {{ChatGPT}} and {UML},'' \emph{SoSym}, pp. 1--13, 2023.

\bibitem{chen2023automated}
K.~Chen, Y.~Yang, B.~Chen, J.~A.~H. L{\'o}pez, G.~Mussbacher, and D.~Varr{\'o},
  ``Automated domain modeling with large language models: A comparative
  study,'' in \emph{MODELS'23}.\hskip 1em plus 0.5em minus 0.4em\relax IEEE,
  2023, pp. 162--172.

\bibitem{stol2020guidelines}
K.-J. Stol and B.~Fitzgerald, ``Guidelines for conducting software engineering
  research,'' in \emph{Contemporary Empirical Methods in Software
  Engineering}.\hskip 1em plus 0.5em minus 0.4em\relax Springer, 2020, pp.
  27--62.

\bibitem{guest2011applied}
G.~Guest, K.~M. MacQueen, and E.~E. Namey, \emph{Applied thematic
  analysis}.\hskip 1em plus 0.5em minus 0.4em\relax sage publications, 2011.

\bibitem{zhao2021natural}
L.~Zhao, W.~Alhoshan, A.~Ferrari, K.~J. Letsholo, M.~A. Ajagbe, E.-V. Chioasca,
  and R.~T. Batista-Navarro, ``Natural language processing for requirements
  engineering: A systematic mapping study,'' \emph{ACM Computing Surveys},
  vol.~54, no.~3, pp. 1--41, 2021.

\bibitem{kof2007scenarios}
L.~Kof, ``Scenarios: Identifying missing objects and actions by means of
  computational linguistics,'' in \emph{RE'07}.\hskip 1em plus 0.5em minus
  0.4em\relax IEEE, 2007, pp. 121--130.

\bibitem{min2023recent}
B.~Min, H.~Ross, E.~Sulem, A.~P.~B. Veyseh, T.~H. Nguyen, O.~Sainz, E.~Agirre,
  I.~Heintz, and D.~Roth, ``Recent advances in natural language processing via
  large pre-trained language models: A survey,'' \emph{ACM Computing Surveys},
  vol.~56, no.~2, pp. 1--40, 2023.

\bibitem{fan2023large}
A.~Fan, B.~Gokkaya, M.~Harman, M.~Lyubarskiy, S.~Sengupta, S.~Yoo, and J.~M.
  Zhang, ``Large language models for software engineering: Survey and open
  problems,'' \emph{arXiv preprint arXiv:2310.03533}, 2023.

\bibitem{hou2023large}
X.~Hou, Y.~Zhao, Y.~Liu, Z.~Yang, K.~Wang, L.~Li, X.~Luo, D.~Lo, J.~Grundy, and
  H.~Wang, ``Large language models for software engineering: A systematic
  literature review,'' \emph{arXiv preprint arXiv:2308.10620}, 2023.

\bibitem{jain2023transformer}
C.~Jain, P.~R. Anish, A.~Singh, and S.~Ghaisas, ``A transformer-based approach
  for abstractive summarization of requirements from obligations in software
  engineering contracts,'' in \emph{RE'23}.\hskip 1em plus 0.5em minus
  0.4em\relax IEEE, 2023, pp. 169--179.

\bibitem{rodriguez2023prompts}
A.~D. Rodriguez, K.~R. Dearstyne, and J.~Cleland-Huang, ``Prompts matter:
  Insights and strategies for prompt engineering in automated software
  traceability,'' in \emph{REW'23}.\hskip 1em plus 0.5em minus 0.4em\relax
  IEEE, 2023, pp. 455--464.

\bibitem{ronanki2023requirements}
K.~Ronanki, B.~Cabrero-Daniel, J.~Horkoff, and C.~Berger, ``Requirements
  engineering using generative {AI}: Prompts and prompting patterns,''
  \emph{arXiv preprint arXiv:2311.03832}, 2023.

\bibitem{white2023chatgpt}
J.~White, S.~Hays, Q.~Fu, J.~Spencer-Smith, and D.~C. Schmidt, ``{ChatGPT}
  prompt patterns for improving code quality, refactoring, requirements
  elicitation, and software design,'' \emph{arXiv preprint arXiv:2303.07839},
  2023.

\bibitem{kundu2013automatic}
D.~Kundu, D.~Samanta, and R.~Mall, ``Automatic code generation from unified
  modelling language sequence diagrams,'' \emph{IET Software}, vol.~7, no.~1,
  pp. 12--28, 2013.

\bibitem{ferrari2017pure}
A.~Ferrari, G.~O. Spagnolo, and S.~Gnesi, ``Pure: A dataset of public
  requirements documents,'' in \emph{RE'17}.\hskip 1em plus 0.5em minus
  0.4em\relax IEEE, 2017, pp. 502--505.

\bibitem{dalpiaz2020conceptualizing}
F.~Dalpiaz and A.~Sturm, ``Conceptualizing requirements using user stories and
  use cases: a controlled experiment,'' in \emph{REFSQ'20}.\hskip 1em plus
  0.5em minus 0.4em\relax Springer, 2020, pp. 221--238.

\bibitem{braun2021saturate}
V.~Braun and V.~Clarke, ``To saturate or not to saturate? {Questioning} data
  saturation as a useful concept for thematic analysis and sample-size
  rationales,'' \emph{Qualitative research in sport, exercise and health},
  vol.~13, no.~2, pp. 201--216, 2021.

\bibitem{white2023prompt}
J.~White, Q.~Fu, S.~Hays, M.~Sandborn, C.~Olea, H.~Gilbert, A.~Elnashar,
  J.~Spencer-Smith, and D.~C. Schmidt, ``A prompt pattern catalog to enhance
  prompt engineering with {{ChatGPT}},'' \emph{arXiv:2302.11382}, 2023.

\bibitem{zowghi2003interplay}
D.~Zowghi and V.~Gervasi, ``On the interplay between consistency, completeness,
  and correctness in requirements evolution,'' \emph{IST}, vol.~45, no.~14, pp.
  993--1009, 2003.

\bibitem{8559686}
``{ISO/IEC/IEEE} international standard - systems and software engineering --
  life cycle processes -- requirements engineering,'' \emph{{ISO/IEC/IEEE
  29148:2018(E)}}, pp. 1--104, 2018.

\bibitem{replicationpkg}
\BIBentryALTinterwordspacing
F.~Alessio, A.~Sallam, and A.~Chetan, ``{Model Generation from Requirements
  with {LLMs}: an Exploratory Study - Replication Package},'' Apr. 2024.
  [Online]. Available: \url{https://doi.org/10.5281/zenodo.10579731}
\BIBentrySTDinterwordspacing

\bibitem{Cohen:60}
J.~Cohen, ``A coefficient of agreement for nominal scales,'' \emph{Educational
  and Psychological Measurement}, vol.~20, no.~1, 1960.

\bibitem{abrahao2011evaluating}
S.~Abrah{\~a}o, E.~Insfran, J.~A. Cars{\'\i}, and M.~Genero, ``Evaluating
  requirements modeling methods based on user perceptions: A family of
  experiments,'' \emph{Inf. Sci.}, vol. 181, no.~16, pp. 3356--3378, 2011.

\bibitem{cohen2013statistical}
J.~Cohen, \emph{Statistical power analysis for the behavioral sciences}.\hskip
  1em plus 0.5em minus 0.4em\relax Academic press, 2013.

\bibitem{clarke2017thematic}
V.~Clarke and V.~Braun, ``Thematic analysis,'' \emph{The journal of positive
  psychology}, vol.~12, no.~3, pp. 297--298, 2017.

\bibitem{ronanki2023investigating}
K.~Ronanki, C.~Berger, and J.~Horkoff, ``Investigating {ChatGPT}’s potential
  to assist in requirements elicitation processes,'' in \emph{SEAA'23}.\hskip
  1em plus 0.5em minus 0.4em\relax IEEE, 2023, pp. 354--361.

\bibitem{sallou2023breaking}
J.~Sallou, T.~Durieux, and A.~Panichella, ``Breaking the silence: the threats
  of using {{LLMs}} in software engineering,'' \emph{arXiv:2312.08055}, 2023.

\end{thebibliography}
